%% file: CTRL_ALT_LED-COMPSAC-final_D.tex
\documentclass[10pt, conference, compsocconf]{IEEEtran}
 \usepackage{graphicx}
\usepackage[table,xcdraw]{xcolor}

\usepackage[cmex10]{amsmath}
\usepackage{amssymb}
\usepackage{cite}
\usepackage{xurl}
\usepackage{algpseudocode}
\usepackage{algorithm}
\usepackage{hhline}
\usepackage{microtype}

\usepackage[subrefformat=parens,labelformat=parens,caption=false,font=footnotesize]{subfig}

\usepackage{caption}
\usepackage{booktabs,tabularx,ragged2e}

\usepackage{svg}
\usepackage{subfig}
\usepackage{float}
\usepackage{adjustbox}
\usepackage{tabularx}
\usepackage{xcolor}
\usepackage{balance}

\usepackage{array,booktabs}
\newcolumntype{L}[1]{>{\raggedright\let\newline\\\arraybackslash\hspace{0pt}}m{#1}}
\newcolumntype{C}[1]{>{\centering\let\newline\\\arraybackslash\hspace{0pt}}m{#1}}
\newcolumntype{R}[1]{>{\raggedleft\let\newline\\\arraybackslash\hspace{0pt}}m{#1}}

\usepackage[hidelinks]{hyperref}
\begin{document}
%
\title{\textbf{CTRL-ALT-\color{red}L\color{green}E\color{yellow}D\color{black}}: Leaking Data from Air-Gapped Computers via Keyboard LEDs }

\author{\IEEEauthorblockN{Mordechai Guri, Boris Zadov, Dima Bykhovsky, Yuval Elovici}
	\IEEEauthorblockA{
		Department of Software and Information Systems 
		Engineering\\Cyber-Security Research Center, Ben-Gurion University of 
		the Negev, Israel\\ 
		\href{mailto:gurim@post.bgu.ac.il}{gurim@post.bgu.ac.il}, 
		\href{mailto:borisza@gmail.com}{borisza@gmail.com}, 
		\href{mailto:bykhov@post.bgu.ac.il}{bykhov@post.bgu.ac.il},
		\href{mailto:elovici@bgu.ac.il}{elovici@bgu.ac.il}\\Air-gap research 
		page: 
		\url{https://cyber.bgu.ac.il/advanced-cyber/airgap}\\ Demo video: 
		\url{https://youtu.be/1kBGDHVr7x0}}}

\maketitle

\begin{abstract}
Using the keyboard LEDs to send data optically was proposed in 2002 by Loughry and Umphress \cite{loughry2002information} (Appendix A).  
In this paper we extensively explore this threat in the context of a \textit{modern} cyber-attack with current hardware and optical equipment. In this type of attack, an advanced persistent threat (APT) uses the keyboard LEDs (Caps-Lock, Num-Lock and Scroll-Lock) to encode information and exfiltrate data from air-gapped computers optically.
Notably, this exfiltration channel is not monitored by existing data leakage prevention (DLP) systems. We examine this attack and its boundaries for today's keyboards with USB controllers and sensitive optical sensors. We also introduce smartphone and smartwatch cameras as components of malicious insider and 'evil maid' attacks.  We provide the necessary scientific background on optical communication and the characteristics of modern USB keyboards at the hardware and software level, and present a transmission protocol and modulation schemes. We implement the exfiltration malware, discuss its design and implementation issues, and evaluate it with different types of keyboards. We also test various receivers, including light sensors, remote cameras, 'extreme' cameras, security cameras, and smartphone cameras. Our experiment shows that data can be leaked from air-gapped computers via the keyboard LEDs at a maximum bit rate of 3000 bit/sec per LED given a light sensor as a receiver, and more than 120 bit/sec if smartphones are used. The attack doesn't require any modification of the keyboard at hardware or firmware levels. 
\end{abstract}

%

\section{Introduction}
In the past decade it has been shown than even air-gapped networks are not immune to breaches. Attackers have used complex attack vectors, such as supply chain attacks and social engineering to compromise air-gapped systems. For example, ten years ago a classified network of the United States military was compromised by a computer worm via a supply chain attack. According to the reports, a foreign intelligence agency supplied infected thumb drives to retail kiosks near NATO headquarters in Kabul. The malicious thumb drive was put into a USB port of a laptop computer that was attached to United States Central Command. The worm spread further to both classified and unclassified networks \cite{grant2009cyber}. 

\subsection{Air-Gap Exfiltration}
Having a foothold in an air-gapped network, the attacker may want to leak information such as files, encryption keys, keylogging information, and so on. Such  behavior is commonly used by espionage malware. However, the \textit{exfiltration} of data from systems with no Internet connectivity is not a trivial task. Over the years,  various communication channels have been developed by researcher which allow attackers to leak data from network-less computers. Using electromagnetic radiation to maintain covert communication has been studied for at least two decades. In this method, a malware controls the electromagnetic emission from a computer and modulates data on top of it. It also have been shown that attackers can exfiltrate data from air-gapped computers using ultrasound, magnetic signals, and even heat emission \cite{guri2016fansmitter,guri2014airhopper,Guri:2018:BAM:3200906.3177230,guri2018lcd}.

In this paper, we examine the threat of leaking data from air-gapped networks via the keyboard LEDs in a modern cyber-attack. We discuss adversarial attack models, and present  design and implementation details. We test a set of USB keyboards and evaluate the use of smartphone cameras and optical sensors as  receivers in the attack. In addition, we evaluate various types of cameras, including remote cameras, 'extreme' cameras, and security cameras. 

The remainder of the paper is organized as follows.  In Section \ref{sec:related} we present related work. Section \ref{sec:Attack_model} describes the adversarial attack model. We provide technical background in Section \ref{sec:TECHNICAL BACKGROUND}. The communication is discussed in Section \ref{sec:comm}, and the implementation is described in Section \ref{sec:imp}. Section \ref{sec:eval} presents the evaluation and results. Countermeasures are discussed in Section \ref{sec:cnt}, and we present our conclusions in Section \ref{sec:conclusion}.

\input{RelatedWork}
\input{ADVERSARIAL_ATTACK_MODEL}
\input{TECHNICAL_BACKGROUND}

\input{communication_dima}

\input{IMPLEMENTATION}
\input{EVALUATION}

\section{Countermeasures}
\label{sec:cnt}
Common countermeasures may include policies aimed to restrict the accessibility of sensitive equipment by placing it in classified rooms where only authorized staff may access it. Typically, all types of cameras (including smartphones and smartwatches) are banned from such secured rooms. However, the banning of cameras is not always feasible because the presence of security and surveillance cameras may also serve as a deterrence measure. Another preventive countermeasure is to disable the keyboard LEDs at the circuit level or cover them. These solutions are not always feasible on a wide scale, since they affect user experience and the keyboard functionality. To protect from remote camera eavesdropping, a special window film that prevents optical eavesdropping may be installed; note that this type of countermeasure doesn't protect against insider and 'evil maid' attacks where the camera is located within the room. Another possible countermeasure is video monitoring the room in order to detect hidden signaling patterns from the keyboard LEDs. 

Software countermeasures may include the detection of the presence of malware that triggers the keyboard LED via its HID USB protocol. Such detection can be implemented using an API hooking technique or USB filter driver such as USBFILTER \cite{tian2016making}. However, such a solution can be bypassed by sophisticated malware with rootkit techniques. It is also possible to limit the bandwidth of the covert channel by implementing a low-pass filter (LPF) at the keyboard driver level. In this case, the LPF will limit the maximum frequency that the status LEDs can be switched on or off. For example, by locking the state of the status LEDs for one second after each change.     
\begin{table*}[ht]
	\caption{Countermeasures}
	\label{tab:CNT}
	\centering
	\begin{tabular}{l|l}
		\toprule
		\textbf{Countermeasure}           & \textbf{Remarks}                                                                                \\
		\midrule
		Banning cameras ('zone' approach) & Expensive. Not always a feasible solution.                                                      \\
		Covering the LEDs                 & Affects the user experience and the keyboard functionality.                                          \\
		Disconnecting the LEDs            & Affects the user experience and the keyboard functionality.                                          \\
		Window covering                   & Expensive, Doesn't protect against insider attacks where the camera is located within the room. \\
		LED activity monitoring           & Can be bypassed by malware or requires an external hardware (camera).                           \\
		Signal jamming                    & Can be bypassed by malware.                                                                     \\
		Low-pass filters (LPF)            & Can be bypassed by malware. Doesn't completely prevent the covert channel.                                 \\
		\bottomrule
	\end{tabular}
\end{table*}

Another approach is to interrupt the emitted signals by intentionally invoking random LED blinking. In this way, the optical signal generated by the malicious code will get mixed up with random blinks. Implementing such a noise generator in a software (within the OS) can be bypassed by a malware, while implementing it within the keyboard firmware requires the involvement of OEMs and may also affect the usability of the keyboard LEDs.
The countermeasures are summarized in Table \ref{tab:CNT}.

\section{Conclusion}
In this paper we show how an attacker can use the keyboard status LEDs (Caps-Lock, Num-Lock and Scroll-Lock) to exfiltrate data from air-gapped computers optically. We examine the attack and its boundaries on modern keyboards with HID USB controllers, sensitive optical sensors, and smartphone cameras. We provide the technical background at the hardware and software level, and present modulation schemes and a transmission protocol. We present design and implementation issues and evaluate the covert channel on different types of keyboards. Our experiment shows that data can be leaked from air-gapped computers via the keyboard LEDs at a bit rate of 3000 bit/sec per LED given a light sensor as a receiver, and more than 120 bit/sec if a smartphone camera is used. 
\label{sec:conclusion}

\balance 
\bibliographystyle{ieeetran}
\bibliography{KBD,cyber_dima}

\end{document}

%% file: RelatedWork.tex
\section{Related Work}
\label{sec:related}
Air-gap covert channels can be categorized as electromagnetic, magnetic, acoustic, thermal, and optical channels \cite{Guri:2018:BAM:3200906.3177230}.

\subsection{Electromagnetic, magnetic, acoustic, and thermal}
In electromagnetic covert channels, the emission generated by various hardware components within the computer is used to carry the leaked information. In 2014, Guri et al introduced AirHopper \cite{guri2014airhopper,guri2017bridging}, a malware that exploits the FM radio signals emanating from the video card to leak data to a nearby smartphone receiver. Guri et al also presented GSMem \cite{guri2015gsmem}, a malware that exploits the electromagnetic emission at GSM, UMTS, and LTE frequencies for air gap exfiltration. The data modulated over the emission can be picked by a low level malware residing in the baseband firmware of a nearby mobile phone. The same researchers also introduced USBee \cite{guri2016usbee}, a malware that used the USB data bus to generate electromagnetic signals to transmit data over the air. In 2018 Guri et al presented ODINI \cite{ODINI} and MAGNETO \cite{guri2018magneto}, two attacks that enable the exfiltration of data via magnetic signals generated by the computer CPU cores. The receiver may be a magnetic sensor or a smartphone located near the computer. Notably, these attacks use low frequency magnetic fields which can bypass Faraday shielding. In 2018, Guri et al also presented PowerHammer, a method of leaking data from air-gapped computers through the power lines \cite{2018powerhammer}. 

Hanspach introduced a method called acoustical mesh networks in air, which enables the transmission of data via high frequency sound waves \cite{hanspach2014covert}. Guri et al also presented Fansmitter \cite{guri2016fansmitter} and DiskFiltration \cite{guri2017acoustic}, two methods enabling the exfiltration of data via sound waves, even when the computers are not equipped with speakers or audio hardware. This research showed how to utilize computer fans and hard disk drive actuator arms to generate covert sound signals. In 2018, Guri et al introduced MOSQUITO \cite{guri2018mosquito} malware that covertly turns speakers connected to a PC into a pair of microphones. Using this technique they established so-called \textit{speaker-to-speaker} air-gap communication between two computers in the same room via ultrasonic waves. BitWhisper \cite{guri2015bitwhisper}, presented in 2015, exploits the computer's heat emissions and PC thermal sensors to create a \textit{thermal covert channel} between computers. This method enabled bidirectional covert communication between two adjacent air-gapped computers. 

\subsection{Optical}
Various types of covert channels proposed over the years to leak data through the air-gap. Back in 2002, Loughry and Umphress \cite{loughry2002information} discussed the threat of information leakage from optical emanations. In particular, they showed that LED status indicators on various communication equipment carries a modulated optical signal correlated with information being processed by the device. In Appendix A of \cite{loughry2002information} the authors presented a threat based on using the keyboard LED for data exfiltration and were able to achieve a transmission bit rate of 150 bit/sec. In this work, we examine this threat in the context of an attack on air-gapped computers. We extend the attack model to malicious insiders who carry smartphones or smartwatches. We also evaluate modern keyboards with USB controllers, and test optical sensors as receivers.

In 2017, Guri et al presented a method code-named LED-it-GO \cite{Guri2017}, which enables data leakage from air-gapped networks via the hard drive indicator LED which exists in almost any PC, server, and laptop today. They showed that a malware can indirectly control the hard drive LED at a rate of 5800Hz which exceeds the visual perception capabilities of humans. In 2018, Guri et al demonstrated a malware which can leak data from air-gapped networks via switch and router LEDs \cite{gur2018xled}. Guri et al presented a covert channel for leaking data through air-gaps using IR (Infrared) light and security cameras \cite{guri2019air}. VisiSploit \cite{guri2016optical} is another optical covert channel in which data is leaked through a hidden image projected on an LCD screen. With this method, the 'invisible' QR code that is embedded on the computer screen is obtained by a remote camera and is then reconstructed using basic image processing operations. Guri also showed how to exfiltrate data from air-gapped computers via fast blinking images \cite{guri2019optical}. 

%% file: ADVERSARIAL_ATTACK_MODEL.tex
\section{Adversarial Attack Model}
\label{sec:Attack_model}
As is common with air-gap covert-channels, the adversarial attack model consists of two malicious components: a transmitter and a receiver. 

\subsection{Transmitter}
The transmitter is a desktop computer or server, attached to a keyboard via the USB port, either directly or through a USB hub or KVM. The computer has to be infected with a malware which gathers sensitive data from the user's computer (e.g., keystrokes, password, encryption keys, documents). The infection of the computer can be achieved via sophisticated attack vectors such as supply chain attacks, social engineering techniques, or with hardware with preinstalled malware \cite{Guri:2018:BAM:3200906.3177230}.  At some point defined by the attacker, the malware starts exfiltrating the data of interest.  The transmission is done by blinking the keyboard LEDs according to the modulation and encoding scheme in use.

\subsection{Receiver}
The receiver is a piece of optical equipment which has a line of sight to the keyboard's LED panel. There are several types of equipment that can used for the reception in this attack model. 
The receiver can be a hidden camera that has a line of sight to the transmitting keyboards, a high resolution camera which is located outside the building but positioned so it has a line of sight to the transmitting keyboards, or a video surveillance closed-circuit TV or IP camera positioned in a location where it has a line of sight to the transmitting keyboards. The receiver can also be a smartphone or wearable video camera (e.g., smartwatch) held by a malicious insider who can position him/herself so as to have a line of sight to the transmitting keyboards, a scenario which is known as the ’evil maid’ attack \cite{rutkowska2009evil}. In this paper we also examine an optical sensor capable of sensing the light emitted from the keyboard LEDs. Such sensors are used extensively in VLC (visible light communication) and LED to LED communication \cite{gur2018xled}. Notably, optical sensors are capable of sampling LED signals at high rates, enabling data reception at a higher bandwidth than a typical video camera. 

\begin{figure}[h!]	
	\centering	
	\includegraphics[width=.9\linewidth]{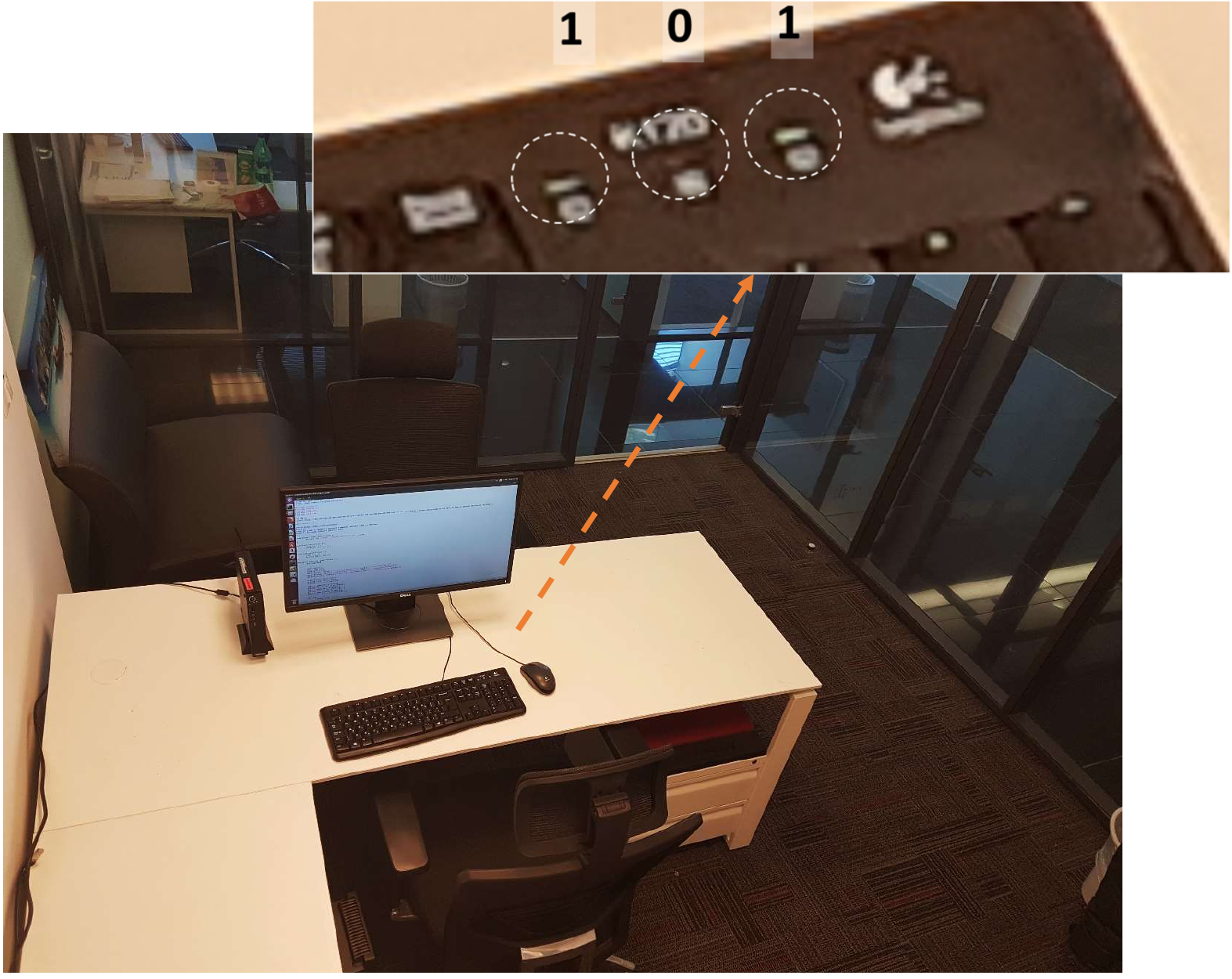}	
	\caption{The binary data is transmitted optically via the keyboard LEDs and recorded by a local camera. In this frame, the binary sequence "101" is encoded.}
	
	\label{fig:example1}
\end{figure}

An illustration of the attack is provided in Figure \ref{fig:example1} in which data is encoded in binary form and covertly transmitted over a stream of LED signals. A compromised security camera films the activity in the room, including the keyboard LEDs. The attacker then applies video processing to decode the signals and reconstruct the modulated data. 

\begin{figure}[h!]
	
	\centering
	
	\includegraphics[width=0.8\linewidth]{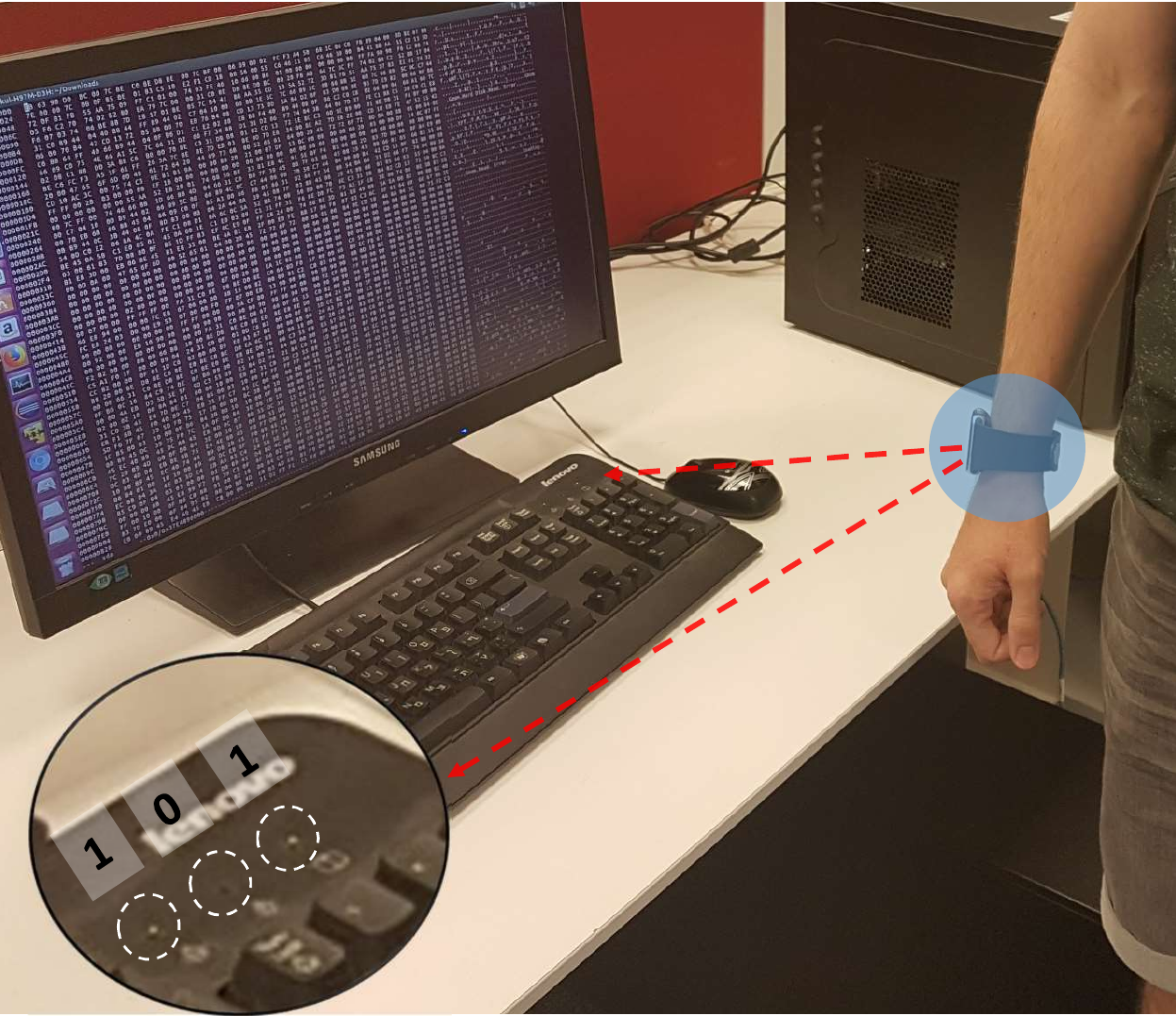}
	
	\caption{An 'evil maid' attack. The binary data is transmitted optically via the keyboard LEDs and recorded by a camera in the smartwatch. In this frame, the binary sequence "101" is encoded.}
	
	\label{fig:example2}
\end{figure}

An illustration of the attack with a malicious insider ('evil maid') scenario is provided in Figure \ref{fig:example2}. In this case, the receiver is a video camera hidden in a smartwatch of a visitor or employee.

%% file: TECHNICAL_BACKGROUND.tex
\section{Technical Background}
\label{sec:TECHNICAL BACKGROUND}
A typical modern PC keyboard contains three toggle keys: Caps-Lock, Num-Lock, and Scroll-Lock. Each key has a corresponding indicator LED, which can be at 'on' or 'off', depending on the state of the lock key. The Num-Lock key was originally used to allow part of the main keyboard to function as a numeric keypad and is rarely in use today. The Caps-Lock causes all letter keys to automatically generate letters in uppercase. The Scroll-Lock was originally used to lock all other scrolling keys. Today the mouse and scroll bars are often used for scrolling, hence the Scroll-Lock is less used.  

Typical users today rarely change the status of the keyboard LED, thus the LEDs can be used by a malware to carry data for exfiltration. Note that many modern keyboards may include additional LEDs and backlights of different colors. In this paper we focus only on the three basic status LEDs which exist in  most consumer keyboards.

\subsection{Status LEDs Controls}
A USB keyboard is a USB HID (human device interface) class device, as defined in the specifications \cite{Microsof71:online}.  Endpoints can be described as data sources or sinks by the USB specifications. The USB HID keyboard initiate IN (input) endpoint object that sends the keystrokes to the host, and OUT (output) endpoint object, that receive the status LEDs settings from the host. At the hardware level, the keyboard consists of a key matrix wired to a micro-controller which in turn recognizes the keystrokes, maps them into corresponding characters, and sends a notification message to the host. The micro-controller also receives output report messages via \textit{Set Report} requests from the host. The requests are used by the host to instruct the micro-controller to change the keyboard LEDs' status \cite{Demonstr51:online}. 

\begin{figure}	
	\centering	
	\includegraphics[page=1,width=\linewidth]{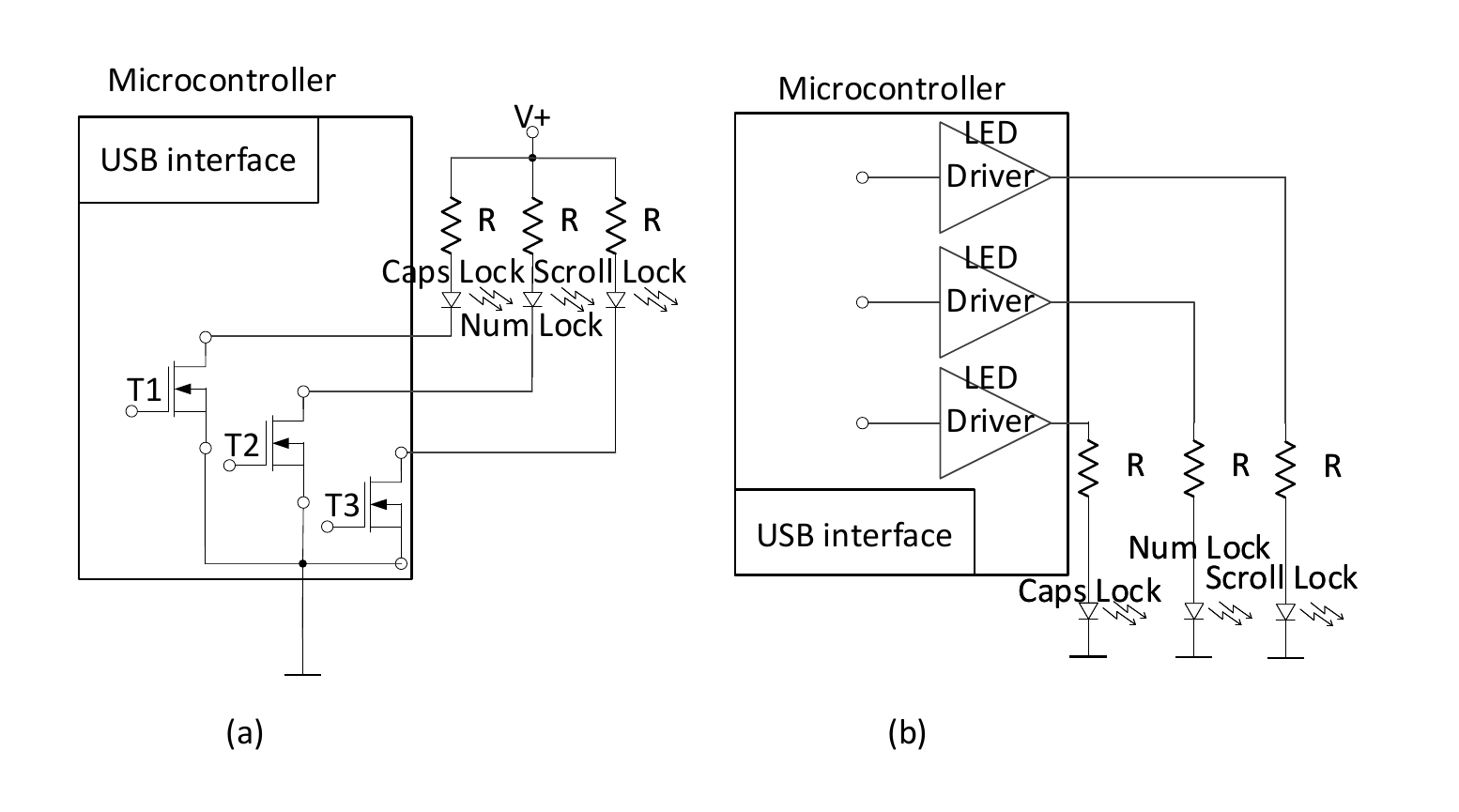}	
	\caption{The implementation of two common keyboard LED driver circuits: (a) MOSFET driver and (b) amplifier.}	
	\label{fig:LED_drivers}
\end{figure}

\subsection{OS Interfaces}
The keyboard LEDs are also exposed to user space processes through the $/sys/class/leds/input$ entries in Linux. The entry $/sys/class/leds/$ contains the properties of each LED, such as name and brightness level (e.g., $numlock/brightness$). Note that most keyboard LEDs don't have hardware brightness support, and hence the brightness value represented by only two states (ON and OFF). The keyboard LED can also be controlled from the Linux kernel by invoking the command \texttt{KDSETLED} of \textit{ioctl()} in the keyboard driver \cite{Flashing65:online}. This approach is preferable in the implementation of a rootkit, in order to evade systems that monitor changes to the keyboard LEDs form the user space. In Windows OS, the \textit{SendInput()} and \textit{keybd\_input()} API functions can be used to control the keyboard LEDs from the user space. Another option is to interact with the USB keyboard programmatically via the HID USB protocol \cite{USBHIDke77:online}. This is done by  sending a request to the device using a standard USB setup transaction defined in the USB Device Class Definition for HIDs \cite{Microsof71:online}.

\subsection{Hardware}
At the hardware level, the circuit in Figure \ref{fig:LED_drivers}(a) is a simple keyboard LED driver based on MOSFET transistor. The MOSFET is used as a power switch where the '1' in the current flows cause the LED to be on. Slightly more advanced circuit in  Figure \ref{fig:LED_drivers}(b) is based on an operational amplifier in comparator configuration with open-loop amplification. It allows faster response and utilizes only two voltage levels (5v and 0v).

%% file: communication_dima.tex
\section{Optical Communication}
\label{sec:comm}
In the section we describe the theory and communication aspects of the proposed covert channel. We also provide a description of the model of the LED based transmitter and outline received optical power.

We discuss this in the context of two types of receivers: an \textit{imaging receiver} (camera) and a \textit{non-imaging receiver} (photo detector sensor). 

\subsection{LED Transmission}
\begin{figure}[h]
	\centering
	\includegraphics[width=.7\linewidth]{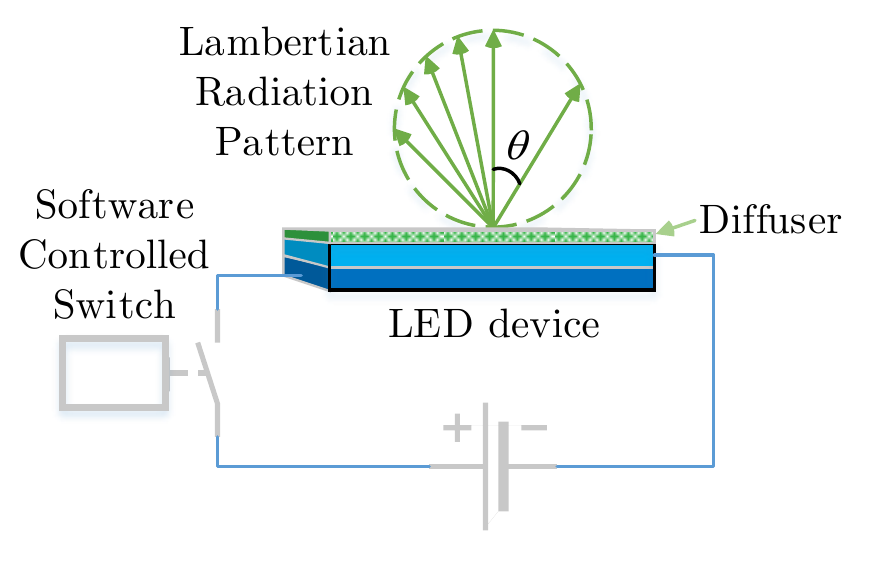}
	\caption{Illustration of Lambertian lighting model.}
	\label{fig:led_transmitter}
\end{figure}

The typical keyboard LED configuration is illustrated in Figure \ref{fig:led_transmitter}. LEDs are typically installed together with a diffuse surface to provide comfortable and homogeneous lighting. The radiation pattern of such a device is modeled by a Lambertian intensity model of the form (measured in steradian\textsuperscript{-1})
\begin{equation}\label{eq:lambertian}
R(\phi )=\frac{1}{\pi}{\cos}\left(\theta\right),
\end{equation}
where $\theta$ is the irradiance angle.

\begin{figure}[h]
	\centering
	\includegraphics[width=.5\linewidth,page=3]{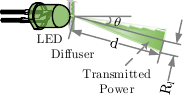}
	\caption{Illustration of the Lambertian lighting model}
	\label{fig:received_power}
\end{figure}

The received optical power is proportional to the solid angle of the receiver (measured in [sr]), calculated by
\begin{equation}
\Omega =\frac{\pi R_{l}^{2}}{d^{2} }
\end{equation}
where $R_l$ is the radius of the outer concentration lens and $d$ is the distance between the LED and the receiver. The relation $R_{l} \ll d$ is assumed. An illustration of the geometric parameters is presented in Figure \ref{fig:received_power}.

Finally, power at the receiver, $P_r$, is calculated by
\begin{equation}
{P}_{r}={P}_{t}\,R(\phi )\,\Omega\,L,
\end{equation}
where $P_t$ is the power of the LED and $L$ is the optical system loss factor.

\subsection{Imaging Receiver (Camera)}
\begin{figure}[h]
\centering
\includegraphics[width=\linewidth,page=5]{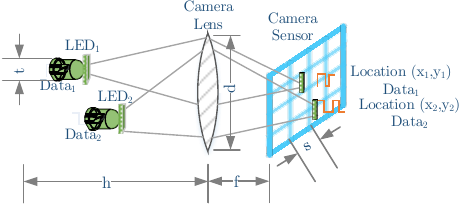}
\caption{Signal acquired by imaging receiver.}
\label{fig:imaging_receiver}
\end{figure}

Cameras can be used to acquire a communication signal. In this case, the signal is focused on a group of sensor pixels, as presented in Figure  \ref{fig:imaging_receiver}. The receiver's performance is limited by two main parameters.
The first parameter is the \textit{diffraction limit} (also referred to as the Rayleigh limit), which constrains the minimum resolvable feature size of the camera and it calculated by \cite{Hecht2016}
\begin{equation}
\tilde{d} \cong 1.22 \frac{\lambda h}{d},
\end{equation}
where $d$ is an aperture size of the camera, $\lambda$ is the wavelength (about 525 nm for green LEDs and 625 nm for red LEDs) and $h$ is the distance to the transmitter (outlined in Figure \ref{fig:imaging_receiver}).  This fundamental limitation is related to wave propagation effects and does not depend on particular camera optics and lenses.

The second parameter is related to the camera magnification. The maximum distance relation for a one pixel imaged object is calculated by \cite{Hecht2016}
\begin{equation}\label{eq:magnification}
\frac{t}{p} = \frac{h}{f},
\end{equation}
where $f$ is the focal distance of the camera, $p$ is a pixel size of a camera array, and $t$ is the size of the transmitting LED.

Multiple LEDs can be used to increase the communication bit rate \cite{Haruyama2015}. The principle of multi-LED communication is illustrated in Figure \ref{fig:imaging_receiver}. Each LED is modulated independently and spatially separated in the camera sensor. 

\subsection{Non-Imaging Receiver}\label{subsec:non_imaging_receiver}
\begin{figure}[h]
	\centering
	\includegraphics[width=.75\linewidth,page=6]{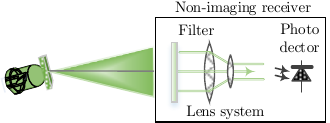}
	\caption{Signal acquired by imaging receiver.}
	\label{fig:non_imaging_receiver}
\end{figure}

The typical receiver includes an appropriate optical filter to reduce the influence of artificial lighting and illumination from the sun. Afterwards, the signal light is concentrated on a photodetector (PD) by an optical lens system, as presented in Figure \ref{fig:non_imaging_receiver}.
The analysis of the performance is similar to that presented in Equations  \eqref{eq:lambertian}-\eqref{eq:magnification}. 

We analyzed the effective distances for a set of basic optical parameters listed in Table \ref{tab:table10}.  
The minimum detectable power level, $P_{thr}$, depends on particular detector parameters and a communication signal frequency \cite{Mackowiak}. For the  parameters applied, the possible communication distance is more than 50 meters. Note, significant axial misalignment may significantly reduce this distance, while the appropriate optical lens system may significantly increase this distance up to an order of magnitude.
\begin{table}[b]
	\centering
	\caption{Evaluation of the effective distance}
	\label{tab:table10}
	\renewcommand{\arraystretch}{1.3}
	\begin{tabularx}{\linewidth}{X|C{1.1cm}|C{1.7cm}|C{1.7cm}}
		\toprule
		\textbf{Parameter}                  & \textbf{Symbol} & \textbf{Value} & \textbf{Typical Range} \\
		\midrule
		Irradiance angle                    &     $\phi$      &   $25^\circ$   &                        \\
		\midrule
		Optical system loss factor          &       $L$       &      0.8       &       0.75-0.95        \\
		\midrule
		Radius of concentration lens        &    $R_{l} $     &  2.54cm (1'')  &       1.5mm-5cm        \\
		\midrule
		Receiver sensitivity (1 kHz signal) &    $P_{thr}$    &      1 nW      &       0.50-2 nW        \\
		\bottomrule
	\end{tabularx}
\end{table}

The comparison between imaging and non-imaging receivers is summarized in Table \ref{tab:comparison_camera_pf}. 
While imaging receivers may be easily implemented by commercial off-the-shelf (COTS) cameras (smartphone camera, webcam, etc.), non-imaging receivers require a dedicated hardware design. The main advantage of non-imaging receivers is their higher communication speeds. 
A high communication range requires accurate axial alignment (pointing), which may be a challenging task for such a receiver.
Moreover, the lower communication speed of imaging receivers may be  compensated somewhat by the adoption of multiple simultaneous transmitters.
\begin{table}
	\centering
	\caption{Comparison of technical characteristics of imaging and non-imaging receivers}
	\label{tab:comparison_camera_pf}
	\renewcommand{\arraystretch}{1.3}	
	\begin{tabular}{l|C{1.8cm}|C{2cm}}
	\toprule
		                         & Imaging Receiver & Non-imaging receiver \\
	\midrule
		Equipment                & COTS             & Custom               \\
		Speed                    & Tens of bps      & kbps                 \\
		Range                    & Low              & Medium-high          \\
		Axial alignment/Pointing & Easy             & Complex              \\
		Parallel communication   & All LEDs         & Single LED           \\
	\bottomrule
	\end{tabular}
\end{table}

%% file: IMPLEMENTATION.tex
\section{Implementation}
\label{sec:imp}
In this section we discuss the data transmission and describe various modulation methods, along with their implementation details. Note that the topic of visible light communication has been widely studied in the last decade. In particular, various modulations and encoding schemes have been proposed for LED to LED communication \cite{komine2004fundamental}. For our purposes, we present basic modulation schemes and describe their characteristics and relevancy to the attack model. As is typical in LED to LED communication, the carrier is the state of the LED, and the basic signal is generated by turning the keyboard LEDs on and off. We denote the two states of an LED (on and off) as LED-ON and LED-OFF, respectively.  We denote the num-lock, caps-lock and scroll-lock keys as $LED_1$, $LED_2$ and $LED_3$ respectively.

\subsection{Malware}

\begin{figure}[h]
	\centering
	\includegraphics[width=\linewidth]{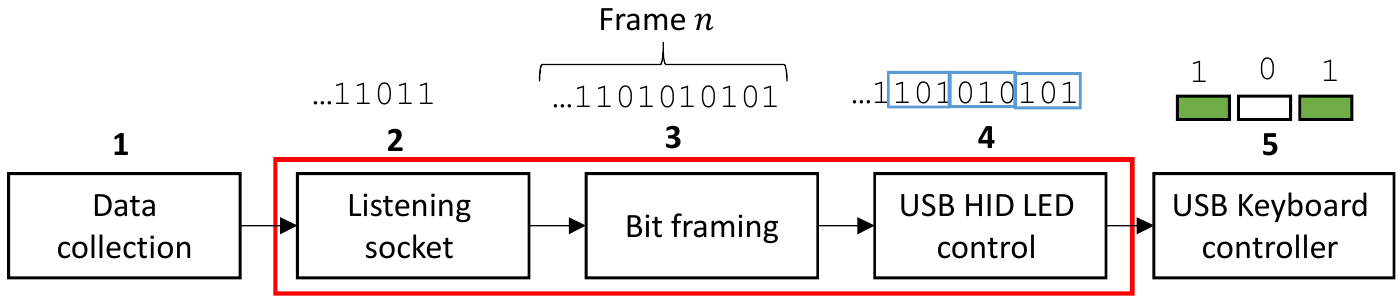}
	\caption{Malware components}
	\label{fig:arch2}
\end{figure}

The malware components are illustrated in Figure \ref{fig:arch2}. The data of interest is collected (1). The data might be keylogging data, encryption keys, passwords, files and so on. The data is then encoded and sent to a listener component (2). The listener component aggregates the data in a form of sequence of bytes. The raw data is  arranged in frames (3) and sent to the modulator (4). The modulator constructs the appropriate HID packets which sent to the USB keyboard controller (5). The packets determine the state of the three LEDs (on/off) given the current three bits.

\begin{figure}[h]
	\centering
	\includegraphics[width=\linewidth]{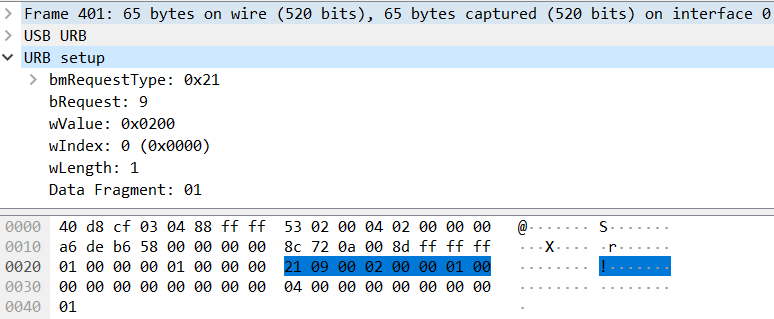}
	\caption{Status LEDs control HID request}
	\label{fig:protocol}
\end{figure}

\subsection{LED Control}
To set the state of the status LEDs (on/off), the module sends a \textit{SetReport} request to the device  with a one-byte data stage Figure \ref{fig:protocol}. The \ packet's request type (\textit{bmRequestType}) is set to 0x21, the request code (\textit{bRequest}) is set to 0x09. The value field of the setup packet (\textit{wValue}) contains the report ID (0x00) in the low byte and the report type (0x02) in the high byte. This indicates a report that is being sent from the software to the hardware. The index field (\textit{wIndex}) contains the interface number of the USB keyboard. The data stage should be 1 byte, which is a bitwise field. When a bit is set to 1, the corresponding LED is turned on. The bits options are specified in Table \ref{tab:bitfield}. Bits 0,1 and 2 determines the status of Num Lock, Caps Lock and Scroll Lock, respectively. The other bits are reserved or used for rarely used LEDs. 

\begin{table}[h]
	\centering
	\caption{LED bits field}
	\label{tab:bitfield}
	\begin{tabular}{l|l}
		\toprule
		\textbf{Bit} & \textbf{Description} \\
		\midrule
		0            & Num Lock             \\
		1            & Caps Lock            \\
		2            & Scroll Lock          \\
		3-7          & Reserved             \\
		\bottomrule
	\end{tabular}
\end{table}
  
\subsection{Data Modulation and Encoding}
We present two single LED modulation schemes: (1) on-off keying (OOK) and (2) binary frequency-shift keying (B-FSK). We also present a scheme which uses all three LEDs to encode data. 

\subsubsection{On-Off Keying (OOK) }
On-off keying is the simplest optical communication modulation. The absence of a signal for a certain duration encodes a logical zero ('0'), while its presence for the same duration encodes a logical one ('1'). In our case, LED-OFF for duration of $T_{off}$  encodes '0' and LED-ON for a duration $T_{on}$ encodes '1.' Note that in the simplest case $T_{on} = T_{off}$. This scheme can use one, two, or three LEDs to modulate data. The theoretical bit-rate for multi-LED communication with OOK modulation is given by
\begin{equation}
R = N \frac{F_r}{2},
\end{equation}
where $N$ is number of the transmitting LEDs and $F_r$ is frame-per-second frequency.

\subsubsection{Binary Frequency-Shift Keying (B-FSK) }
Frequency-shift keying (FSK) is a modulation scheme in which digital information is modulated through a frequency changes in a carrier signal. In the B-FSK only two frequencies, usually representing zero and one, are used for the modulation. In our case, LED-ON for duration of $T_{off}$  encodes a logical zero and LED-ON for a duration $T_{on}$ encodes a logical one. Note that in the simple case $T_{on} = T_{off}$. We make a separation between two sequential bits by setting the LED in the off state for time interval $T_d$. This scheme uses a one, two, or three LEDs to modulate data.

\subsubsection{Amplitude Shift Keying (ASK) - all LEDs}

In this scheme we use three LEDs to represent a series of three bits. As in OOK encoding, the absence of a signal for a certain time duration encodes a logical zero for a specific LED, while its presence for the same time duration encodes a logical one for a specific LED. All of the LEDs remain in the same status for a duration of $T_{all}$ and then change to the next state. This encoding is relevant for cases where several LEDs in the keyboard are available for the transmission. We separate between two sequences of bits by setting the all of the LEDs in the '000' state for time interval $T_d$. The ASK encoding is illustrated in Table \ref{tableask}.

\begin{table}[]
	\renewcommand{\arraystretch}{1.3}
	\caption{ASK modulation with all LEDs}
	\centering
	\label{tableask}
		\begin{tabular}{c|c|c||c|c}
			\hline
			\textbf{$LED_1$} & \textbf{$LED_2$} & \textbf{$LED_3$} & \textbf{Duration} & \textbf{} \\ \hline
			&  &  & $T_{all}$ & 000 \\ \hline
			\cellcolor[HTML]{32CB00}{\color[HTML]{036400} } &  &  & $T_{all}$ & 100 \\ \hline
			& \cellcolor[HTML]{32CB00} &  & $T_{all}$ & 010 \\ \hline
			\cellcolor[HTML]{32CB00} & \cellcolor[HTML]{32CB00} &  & $T_{all}$ & 110 \\ \hline
			&  & \cellcolor[HTML]{32CB00} & $T_{all}$ & 001 \\ \hline
			\cellcolor[HTML]{32CB00} &  & \cellcolor[HTML]{32CB00} & $T_{all}$ & 101 \\ \hline
			\cellcolor[HTML]{32CB00} & \cellcolor[HTML]{32CB00} &  & $T_{all}$ & 110 \\ \hline
			\cellcolor[HTML]{32CB00} & \cellcolor[HTML]{32CB00} & \cellcolor[HTML]{32CB00} & $T_{all}$ & 111 \\ \hline
			&  &  & $T_{d}$ & Separation \\ \hline
		\end{tabular}%
\end{table}

\subsection{Bit Framing}

We transmit the data in small packets called frames. Each frame is composed of a preamble, a payload, and a checksum. The preamble consists of a sequence of eight alternating bits ('10101010') and is used by the receiver to periodically determine the properties of the channel, such as $T_{on}$ and $T_{off}$. In addition, the preamble header allows the receiver to identify the beginning of a transmission and calibrate other parameters, such as the location, intensity and color of the keyboard LEDs. The payload is the raw data to be transmitted. In our case, we arbitrarily choose 256 bits as the payload size. For error detection, we add a CRC (cyclic redundancy check) value, which is calculated on the payload and added to the end of the frame. The receiver calculates the CRC for the received payload, and if it differs from the received CRC, an error is detected. More efficient bit framing may employ variable length frames, error correction codes, and compression, and is beyond the scope of our discussion.

%% file: EVALUATION.tex
\section{Evaluation}
\label{sec:eval}
In this section we evaluate the optical covert channel in terms of distance and bit rate. During the evaluation, we have tested four types of COTS USB keyboards that are listed in Table \ref{table:keybords}. 
\begin{table}[]
\caption{The tested keyboards}
\label{table:keybords}
\centering
\renewcommand{\arraystretch}{1.3}	
		\begin{tabular}{l|l|l}
			\toprule
			\textbf{\#} & \textbf{Vendor} & \textbf{Model} \\
			\midrule
			1           & Dell            & KB212-B        \\
			\midrule
			2           & Lenovo          & SK-8825        \\
			\midrule
			3           & Logitech        & K120           \\
			\midrule
			4           & SilverLine      & MM-KB2011      \\
			\bottomrule
		\end{tabular}%
\end{table}

\subsection{Camera Receivers}
There are two types of receivers relevant to the attack model: cameras and light sensors. Receiving the optical signals by a camera depends on the line of sight and visibility of the keyboard. We process the recorded video in order to detect the location of each transmitting keyboard and its LEDs. The video is processed frame by frame to identify the LED state for each frame. Finally, the binary data is decoded based on the encoding scheme. 

\subsubsection{Video processing}
For decoding the videos we used OpenCV 3.2 \cite{OpenCVli0:online}, which is an open-source computer vision library that focuses on real-time video processing for academic and commercial use. We developed a C program that receives the video as an input and saves each LED's timings and state (illumination amplitude) to an output file. To detect and enumerate LED blinks, we used the fundamental approaches used in LED based communication \cite{komine2004fundamental,vuvcic2010513}. For each LED in the frame, we calculated the brightness function $p_n (x,y)$, where $x$ and $y$ are the coordinates of a pixel in the image, and $n$ is the frame number in the sampled video. Our output is an intensity vector $S_(x,y)(p_0, p_1,…, p_N)$,  which describes the change of pixel intensity in time. The brightness of the LED is a quantized level of light intensity in the image at the point in the 2D space. The algorithm determines the on and off brightness threshold values using the temporal mean of the sampled signal. Based on the intensity vector and threshold values, we demodulate the signals encoded in the video.

As expected, the main factor in determining the maximum bit rate for video cameras is the number of frames per second (FPS). In our experiments, we identified two to three frames per bit as the optimal setting needed to successfully detect the LED transmissions in most cameras. We tested various types of cameras as receivers. All of the transmissions were decoded using the video processing demodulator. Table \ref{table:max_tare} shows the maximal bit rate achieved for each video camera.

\begin{table}[]
	\renewcommand{\arraystretch}{1.3}
	\centering
	\caption{Maximum bit rate of different receivers }
	\label{table:max_tare}
		\begin{tabularx}{\columnwidth}{X|C{10mm}|c|c}
			\toprule
			\textbf{Tested Camera/Sensor}                      & \textbf{Max FPS} & \textbf{OOK/FSK} & \textbf{Three LEDs} \\
			\midrule
			Entry-level DSLR \newline (Nikon D7100)            &        60        &    15 bit/sec    &     45 bit/sec      \\
			\midrule
			High-end security camera \newline (Sony SNC-EB600) &        30        &    15 bit/sec    &     45 bit/sec      \\
			\midrule
			HD Webcam \newline (Microsoft LifeCam)             &        30        &    15 bit/sec    &     45 bit/sec      \\
			\midrule
			Smartphone camera \newline (Samsung Galaxy S7)     &     30 - 120     &  15-45 bit/sec   &   45-130 bit/sec    \\
			\bottomrule
		\end{tabularx}%
\end{table}

\begin{table}[]
	\centering
	\caption{The maximum distance for 30 bit/sec}
	\label{tab:ber1}
	\renewcommand{\arraystretch}{1.3}	
	\begin{tabular}{c|c|c}
		\toprule
		 \textbf{Keyboard}  & \textbf{Distance} & \textbf{Bit rate (OOK) with BER of $\leq$ 1\%} \\
		\midrule
		   \textbf{Dell}    &     0 - 9.5m      &                   30 bit/sec                   \\
		\midrule
		  \textbf{Lenovo}   &     0 - 9.5m      &                   30 bit/sec                   \\
		\midrule
		 \textbf{Logitech}  &     0 - 6.5m      &                   30 bit/sec                   \\
		\midrule
		\textbf{Silverline} &     0 -  9.5m     &                   30 bit/sec                   \\
		\bottomrule
	\end{tabular}%
	
\end{table}

\subsubsection{Smartphone camera distance}
Smartphone cameras might be used in an 'evil maid' attack to record the keyboard LEDs.
We evaluated the practical distances at which the smartphone camera can operate in a practical attack. In particular, we wanted to determine the maximal distance that we could maintain a bit rate of 30 bit/sec with an acceptable BER (bit error rate) of less than 1\%. The results are presented in Table \ref{tab:ber1}. For three of the four keyboards the maximum distance we achieved for a bit rate of 30 bit/sec is 9.5 meters. With the Logitech keyboard we achieved a limited distance of 6.5 meters, mainly due to the low power of it's status LEDs.

\subsection{Light Sensor Receivers}
\begin{figure}[h!]
	\centering
	\includegraphics[page=2,width=\linewidth]{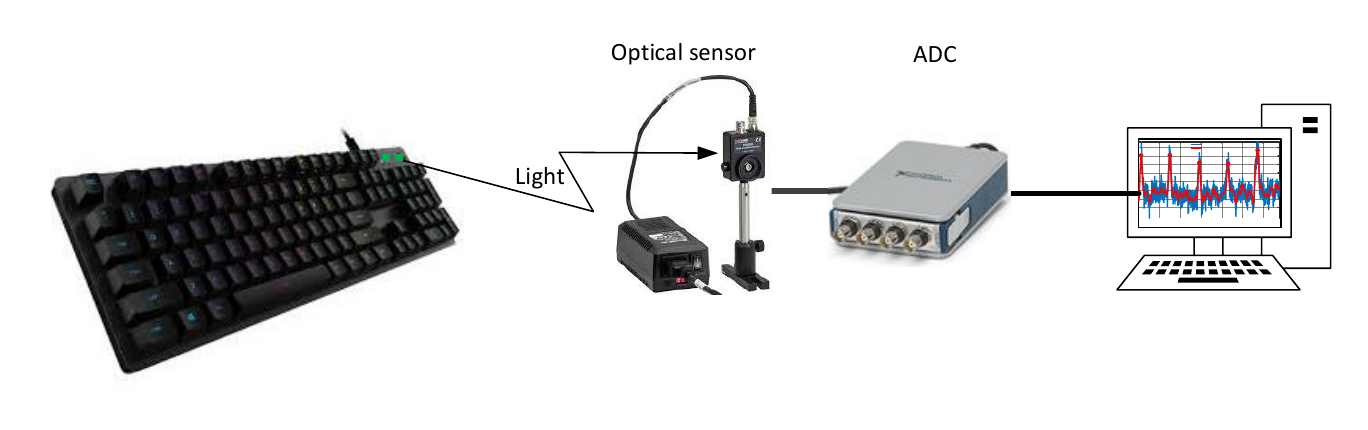}	
	\caption{The measurement setup with the Thorlabs PDA100A light sensor and NI-9233 data acquisition hardware.}	
	\label{fig:measurement_setup}
\end{figure}

A photodiode is a semiconductor that converts light into electrical current. To evaluate the transmissions at high speeds, we built a measurement setup based on photodiode light sensors (Figure \ref{fig:measurement_setup}). The Thorlabs PDA100A light sensor \cite{Thorlabs} is connected to an internal charge amplifier and a data acquisition system. We also used an optical zoom lens to focus on the sensing area and reduce the optical noise. The data is sampled with the National Instruments cDAQ portable sensor measurement system \cite{Ni2} via a 16-bit analog-to-digital NI-9223 card \cite{Ni1} which is capable of 1 Msamples per second. The light emitted from the transmitting keyboard is sampled by the sensor and processed by MATLAB software. 

\subsubsection{Measurement Setup}
\label{sec:measure}
The PDA100A  includes a reverse-biased PIN photodiode, mated to a switchable gain transimpedance amplifier, and packaged in a protective cover. 

The responsivity, $R$, of the photodiode can be defined as a ratio of generated photocurrent, $I_{PD}$, to the incident light power, $P$, at a given wavelength,
\begin{equation}
R = \frac{I_{PD}}{P}.
\label{eq:1}
\end{equation}

The gain of the sensor, $A$, in our measurements is ${4.75 \cdot 10^5 V/A}$, the PDA responsivity of the sensor for green light is $R=0.32$ A/W, and the output voltage is given by
\begin{equation}
V_{out} = P_{in} R A.
\label{eq:2}
\end{equation}

\subsubsection{OOK}
In this experiment we tested the maximal frequency at which the keyboard LEDs can blink when controlled from a user space program or shellcode running within the keyboard's OS. The blinking frequency is important, since it defines the maximum communication speed of the the LED.

\begin{figure}[!ht]
\centering
		\subfloat[]{\includegraphics[width=0.49\columnwidth]{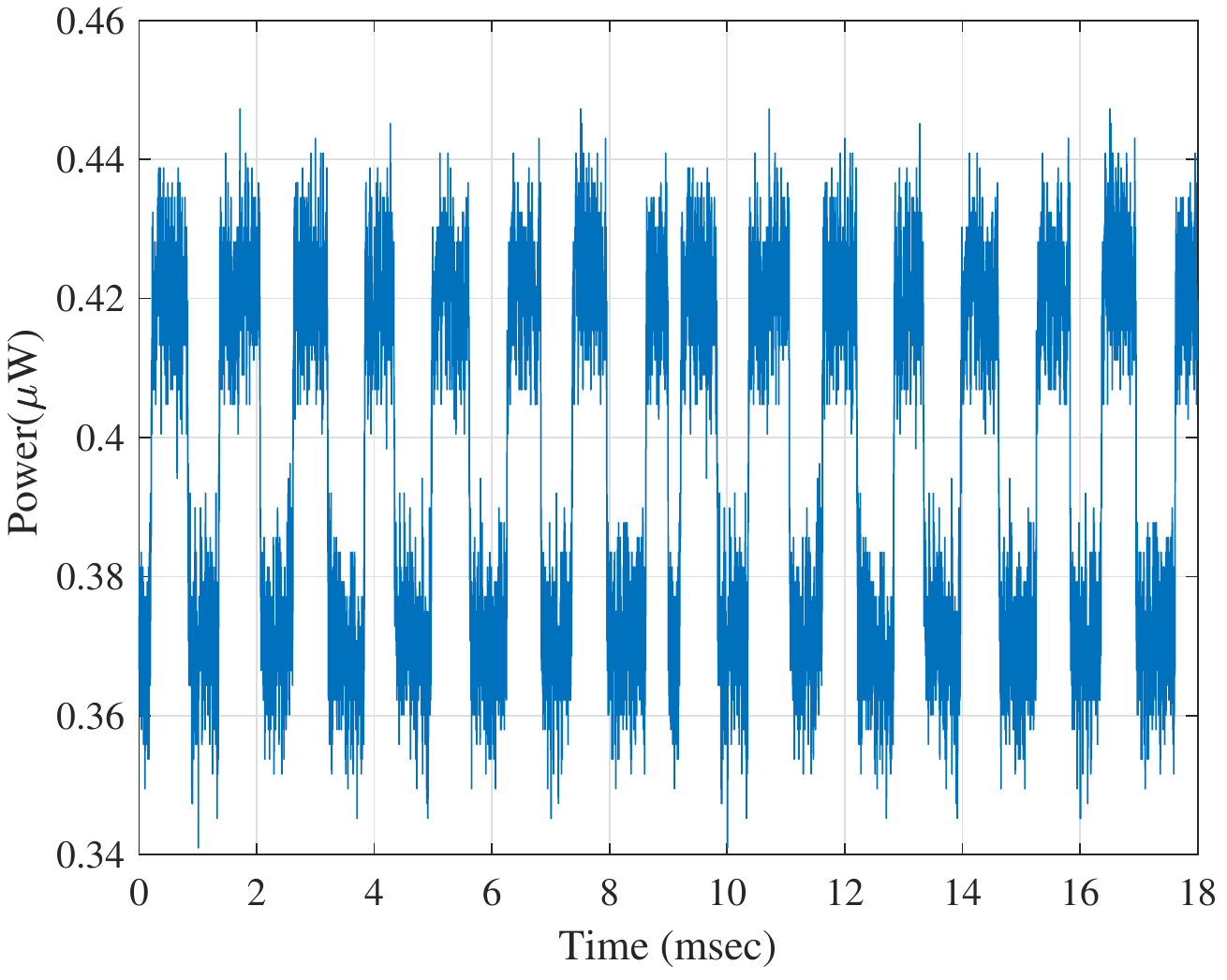}}
		\hfil
		\subfloat[]{\includegraphics[width=0.49\columnwidth]{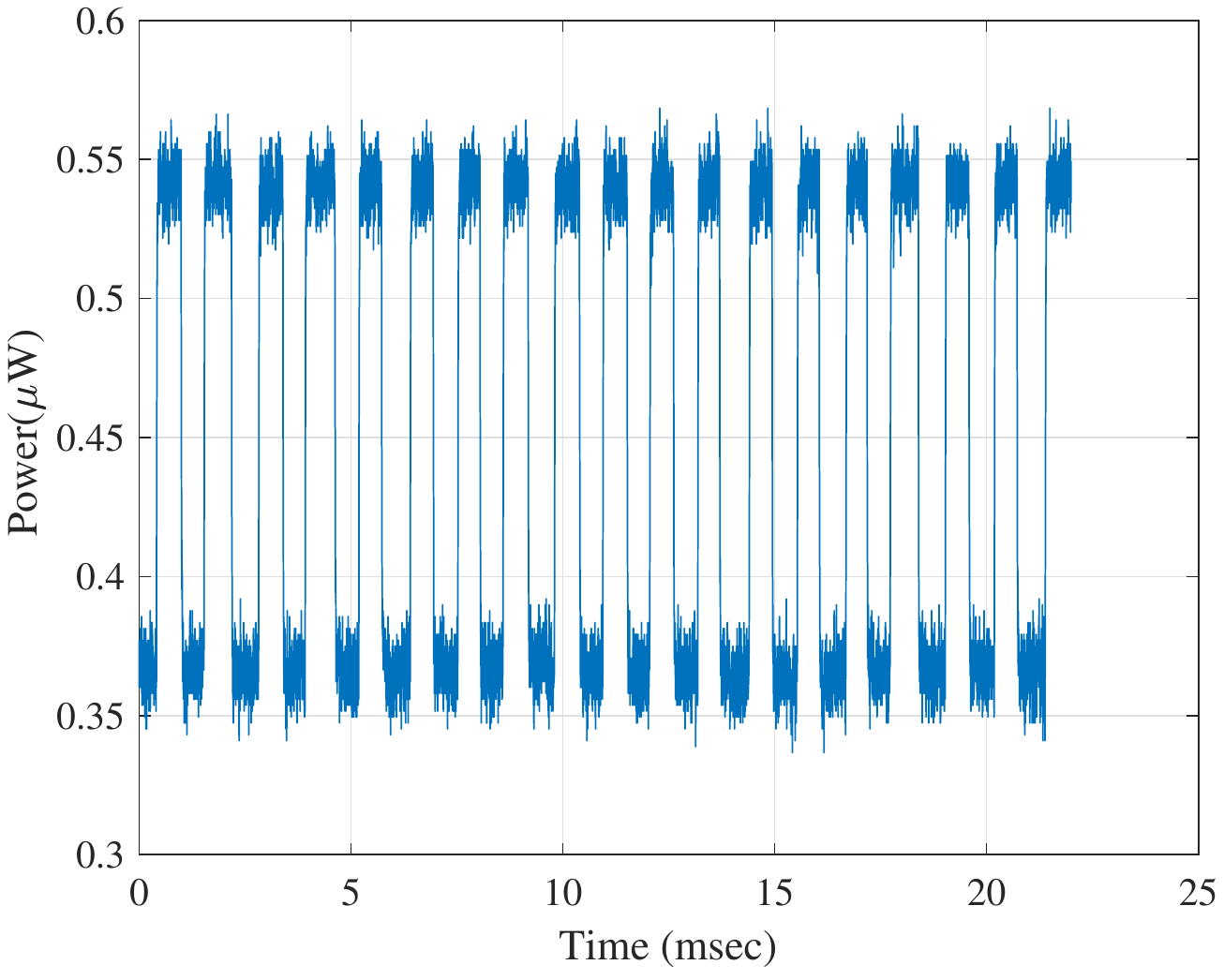}}

		\subfloat[]{\includegraphics[width=0.49\columnwidth]{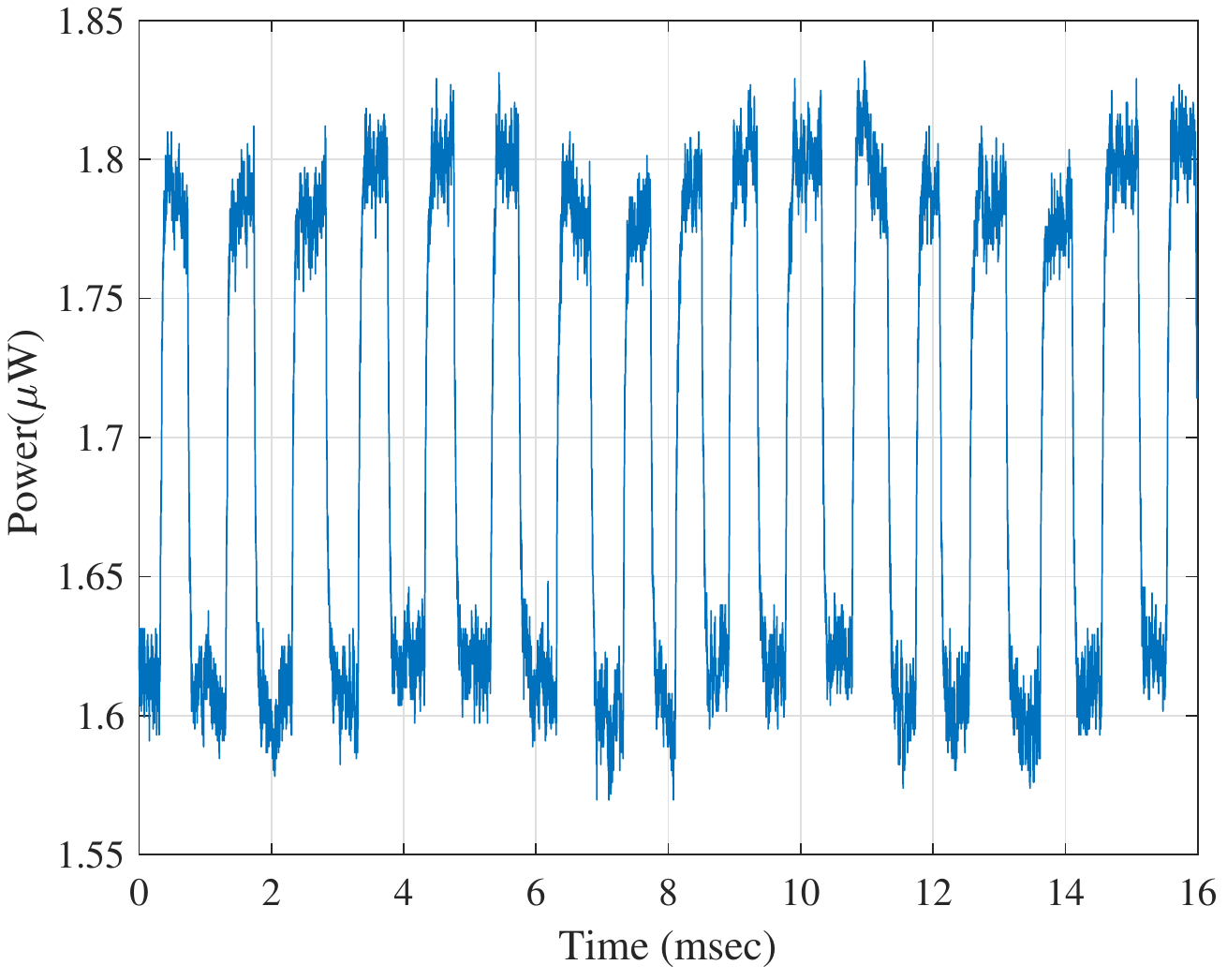}}
				\hfil
		\subfloat[]{\includegraphics[width=0.49\columnwidth]{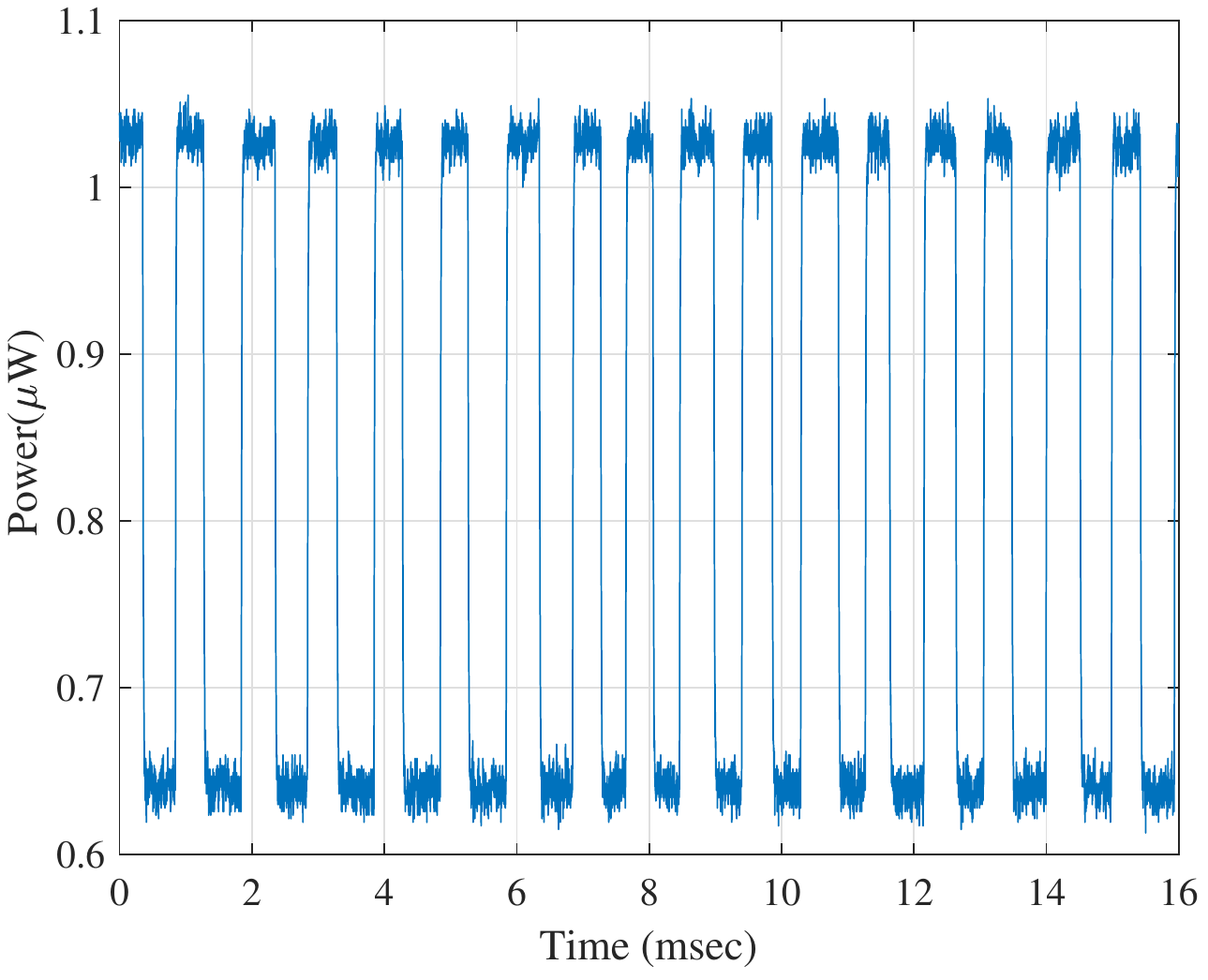}}
		
		\subfloat[]{\includegraphics[width=0.49\columnwidth]{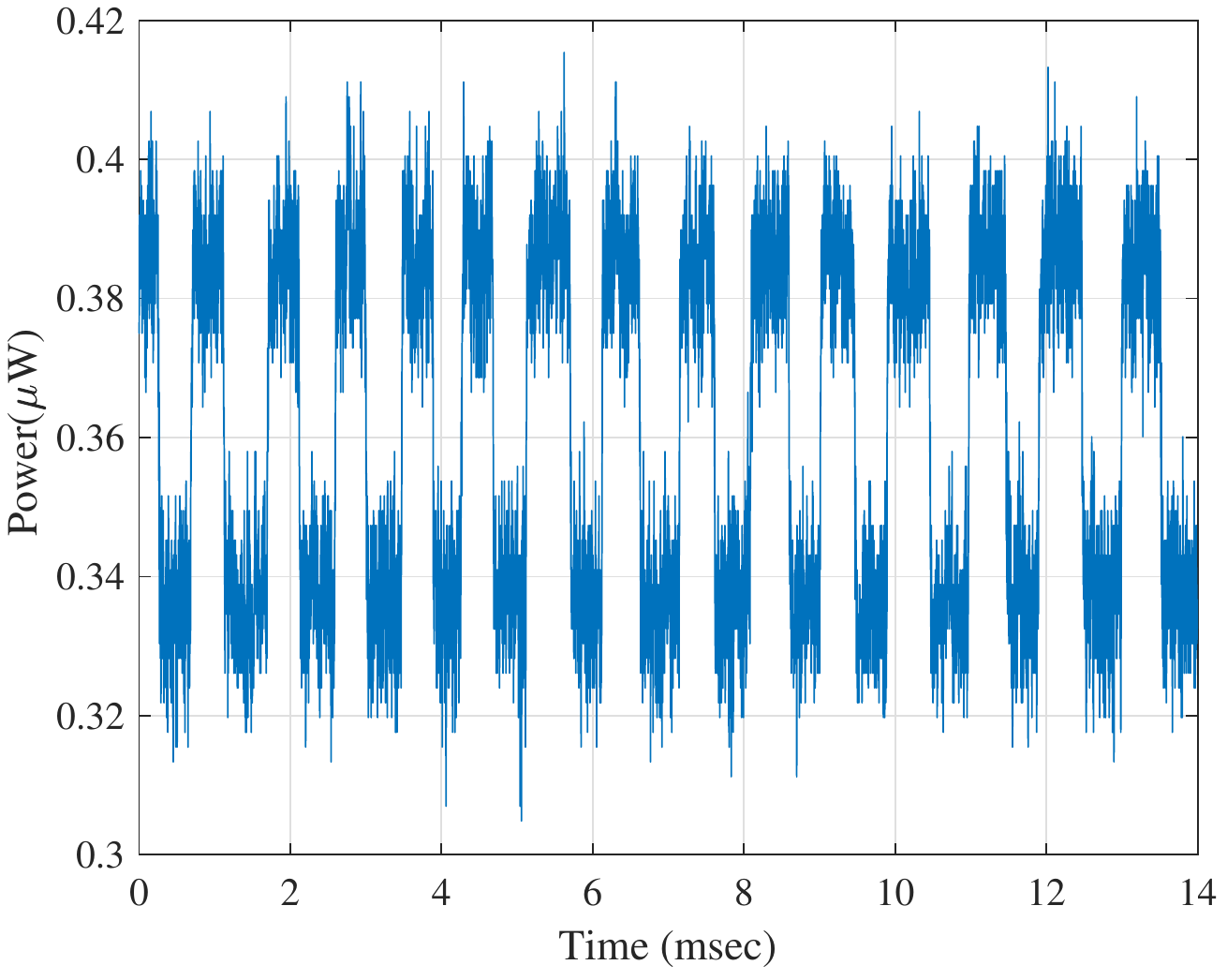}}
		\hfil
		\subfloat[]{\includegraphics[width=0.49\columnwidth]{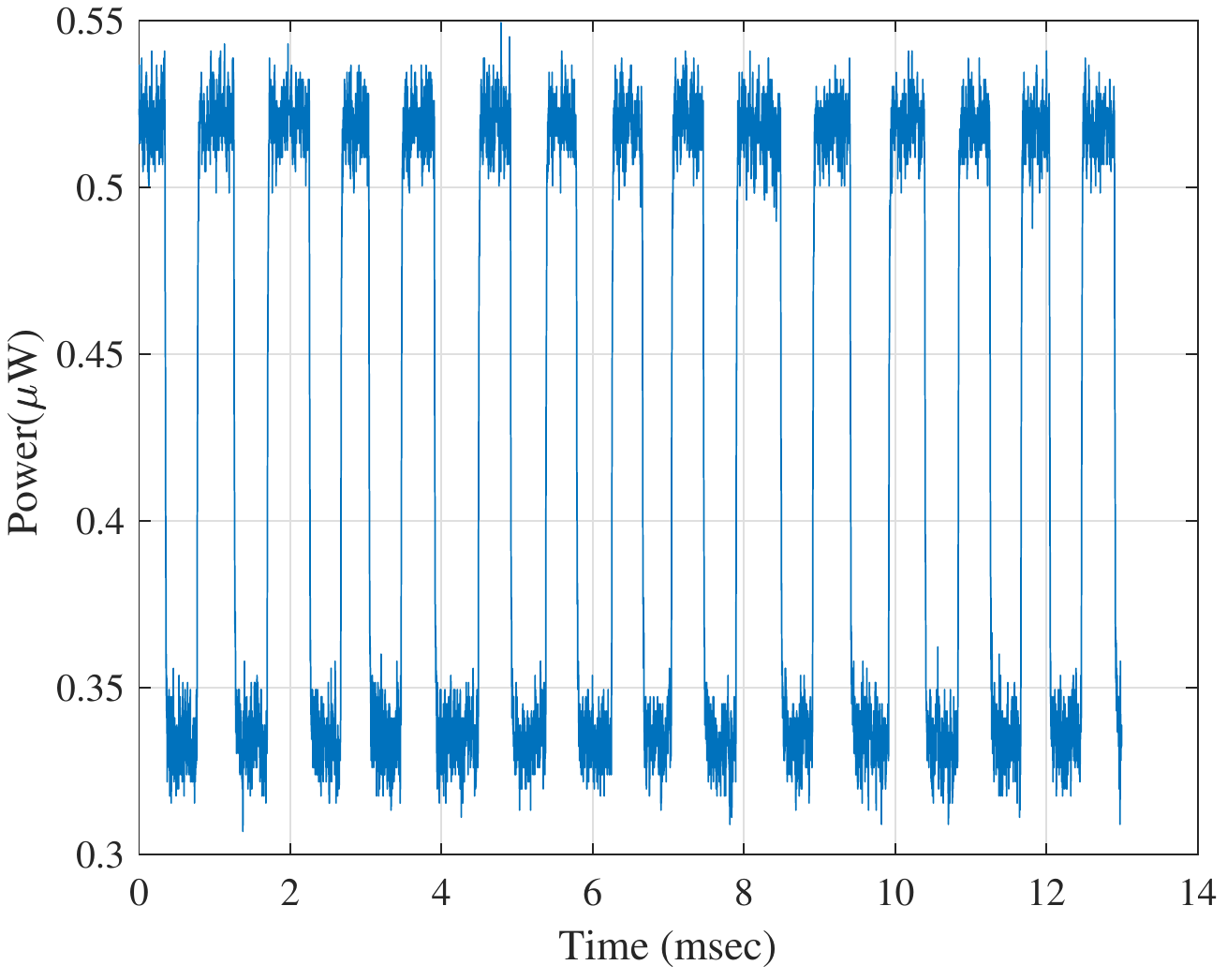}}
		
		\subfloat[]{\includegraphics[width=0.49\columnwidth]{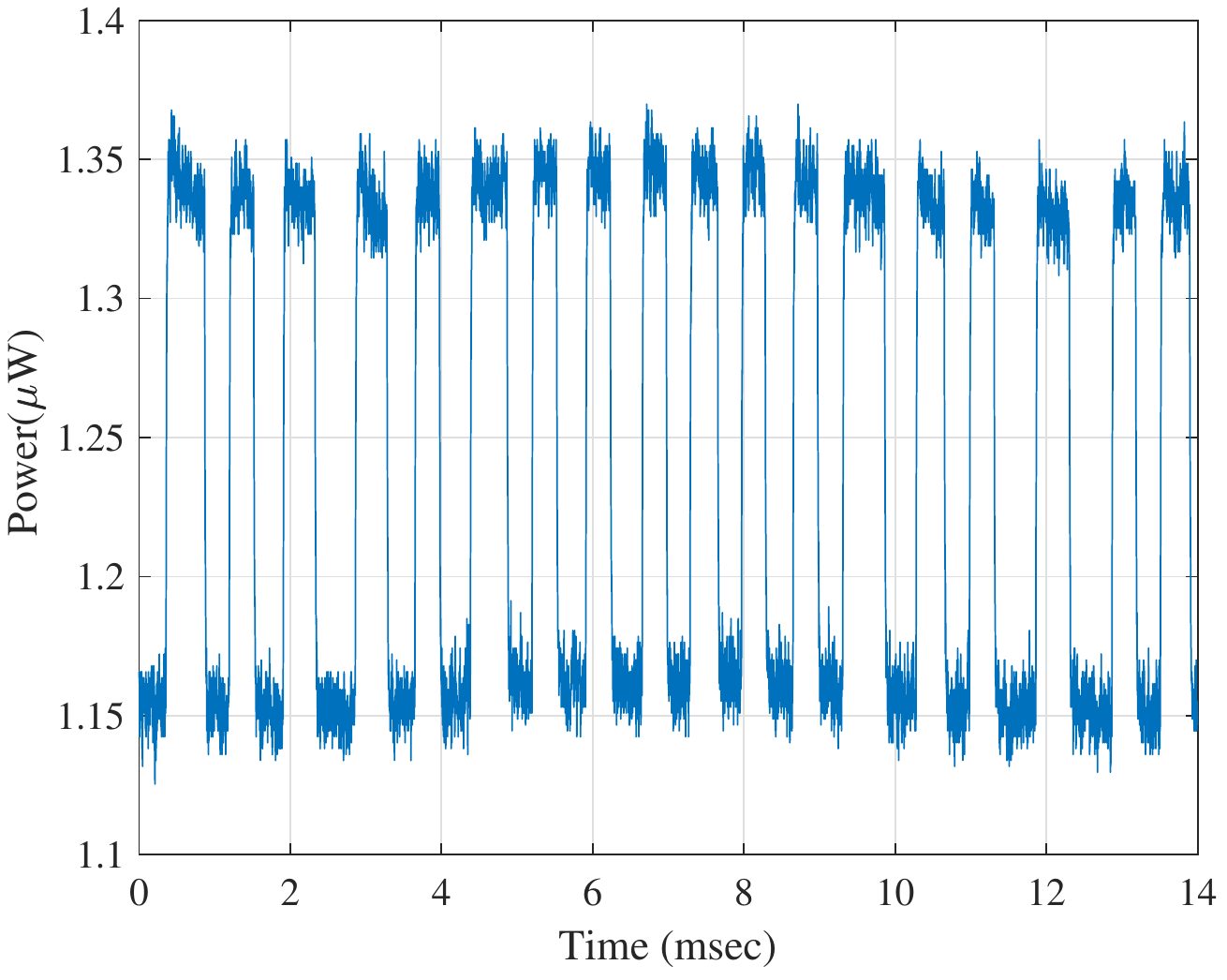}}
			\hfil
		\subfloat[]{\includegraphics[width=0.49\columnwidth]{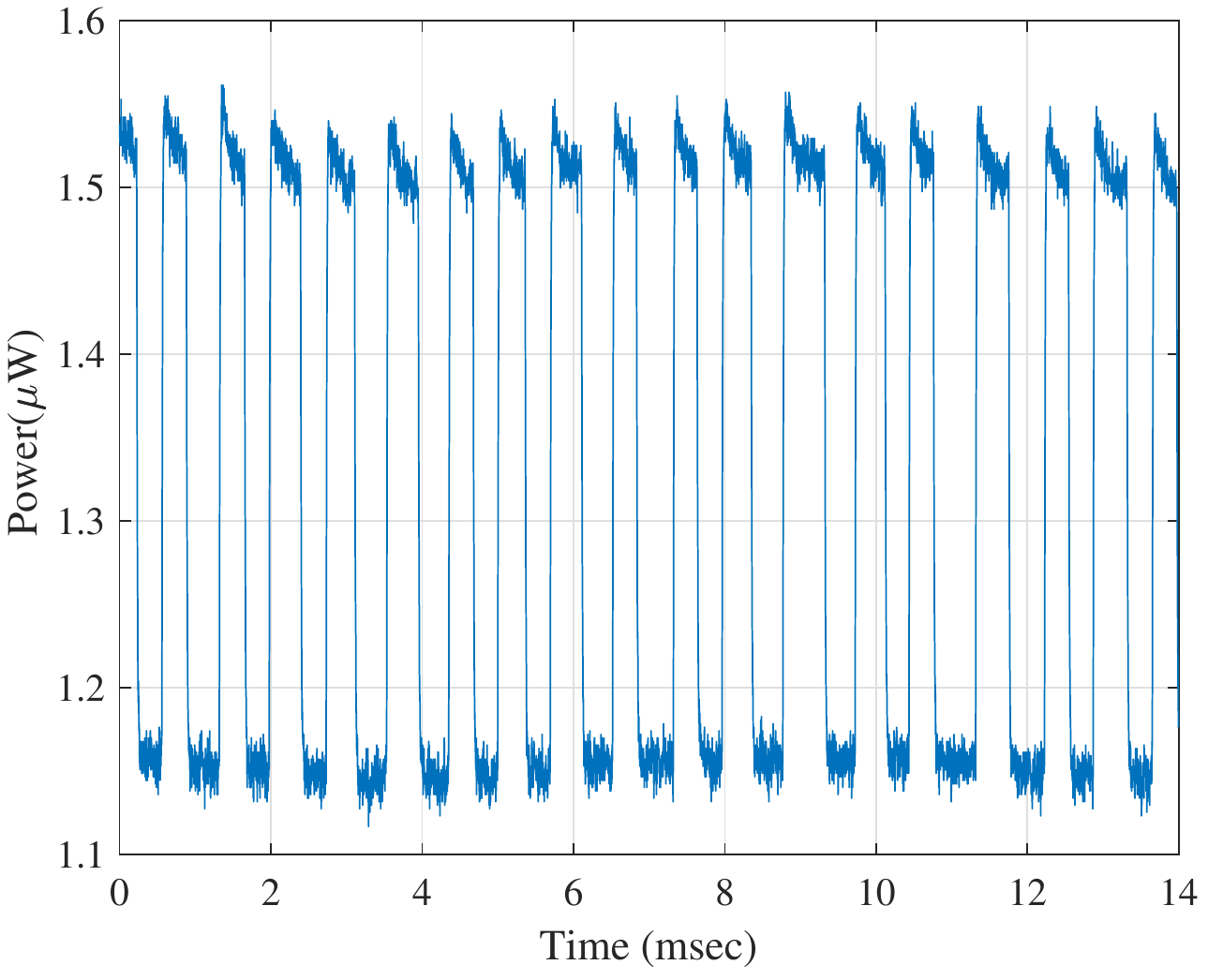}}
			
	\caption{Maximum speed of the basic signal for: (a) Dell 1 LED, (b) Dell 3 LEDs, (c) Lenovo 1 LED,  (d) Lenovo 3 LEDs, (e) Logitech 1 LED,  (f) Logitech 3 LEDs, (g) Silverline 1 LED, (h) SilverLine 3 LEDs.}
	\label{fig:max_speed}
\end{figure}

Figures \ref{fig:max_speed}(a)(c)(e)(g) show the signals as received from different keyboards when its leftmost LED is repeatedly turned on and off. The sampling rate in this test is 500 Ksamples per second. As can be seen, the minimal LED-ON time is approximately 600 $\mu$s for Dell, 440$\mu$s for Lenovo, 400$\mu$s for Logitech, 400$\mu$s for Silverline. The minimal blinking time (LED-ON, LED-OFF) is 800$\mu$s, which implies a bit rate of 1250 bit/sec with the simplest OOK modulation. During the LED-ON time the sampled powers are approximately 0.42mW, 1.8mW, 0.42mW, and 1.35mW respectively, while for LED-OFF powers are 0.37mW, 1.6mW, 0.33mW, and 1.15mW respectively and are resulted by the ambient lighting in the room.

Figures \ref{fig:max_speed}(b)(d)(f)(h) show the signals as received from different keyboards when all three LEDs are repeatedly turned on and off. By using all of the LEDs together for modulation, we have significantly increased the optical signals emitted from the transmitting keyboard. This method can be used when the optical signal level generated by a single LED is too low for successful reception. As can be seen, with multiple LEDs the minimal blinking time (LED-ON, LED-OFF) is approximately 280$\mu$s, 500$\mu$s, 440$\mu$s, and 400$\mu$s respectively, which implies the corresponding bit rates of 3570, 2000, 2270, and 2500 bit/sec with the simplest OOK modulation.

\subsubsection{Multiple LEDs ASK}
With a camera receiver it is possible to distinguish between two or more different transmitting LEDs. In this case the bit rate is derived from the number of LEDs available for modulation. That is, with $N$ LEDs we can generate $2^N$ different signals. Unlike camera receivers, light sensors can only measure the amount of light emitted from the keyboard and cannot distinguish between different LEDs. One straightforward strategy is to use OOK modulation when '0' is modulated with all of the LEDs in the OFF state, and '1' is modulated with all of the LEDs in the ON state. Obviously, this type of modulation limits the transmission rate. We found that under some circumstances it is also possible to distinguish between different amounts of light emitted when using different numbers of LEDs, even with a light sensor. Consequentially, we can increase the bit rate by modulating multiple bits with several LEDs (using ASK modulation) when a light sensor is used for reception. Under optimal conditions $n$ different amplitudes can modulate $log_2(n)$ values.

\begin{figure}[!ht]
\centering
		\subfloat[]{\includegraphics[width=0.49\columnwidth]{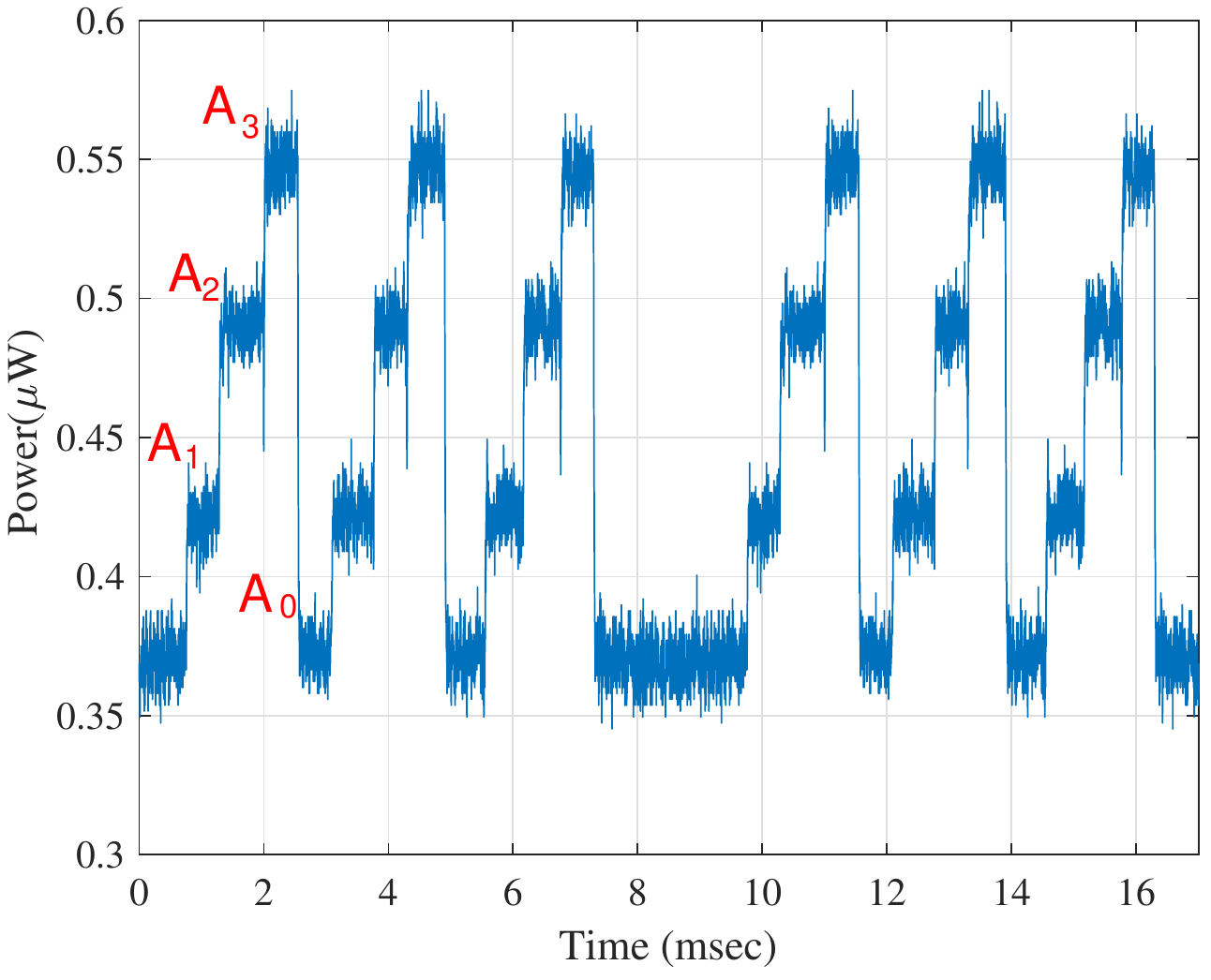}}\hfil
		\subfloat[]{\includegraphics[width=0.49\columnwidth]{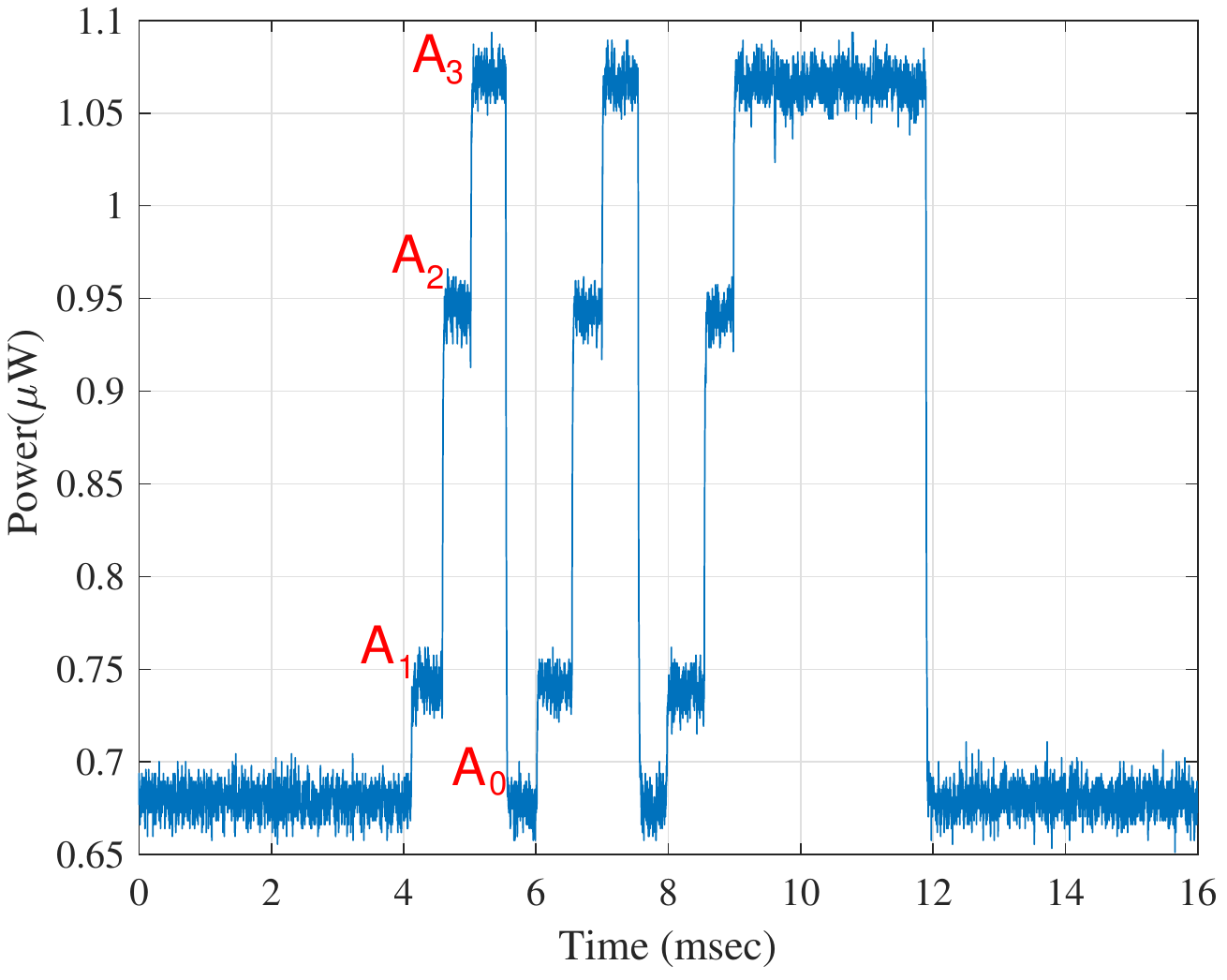}}
				
		\subfloat[]{\includegraphics[width=0.49\columnwidth]{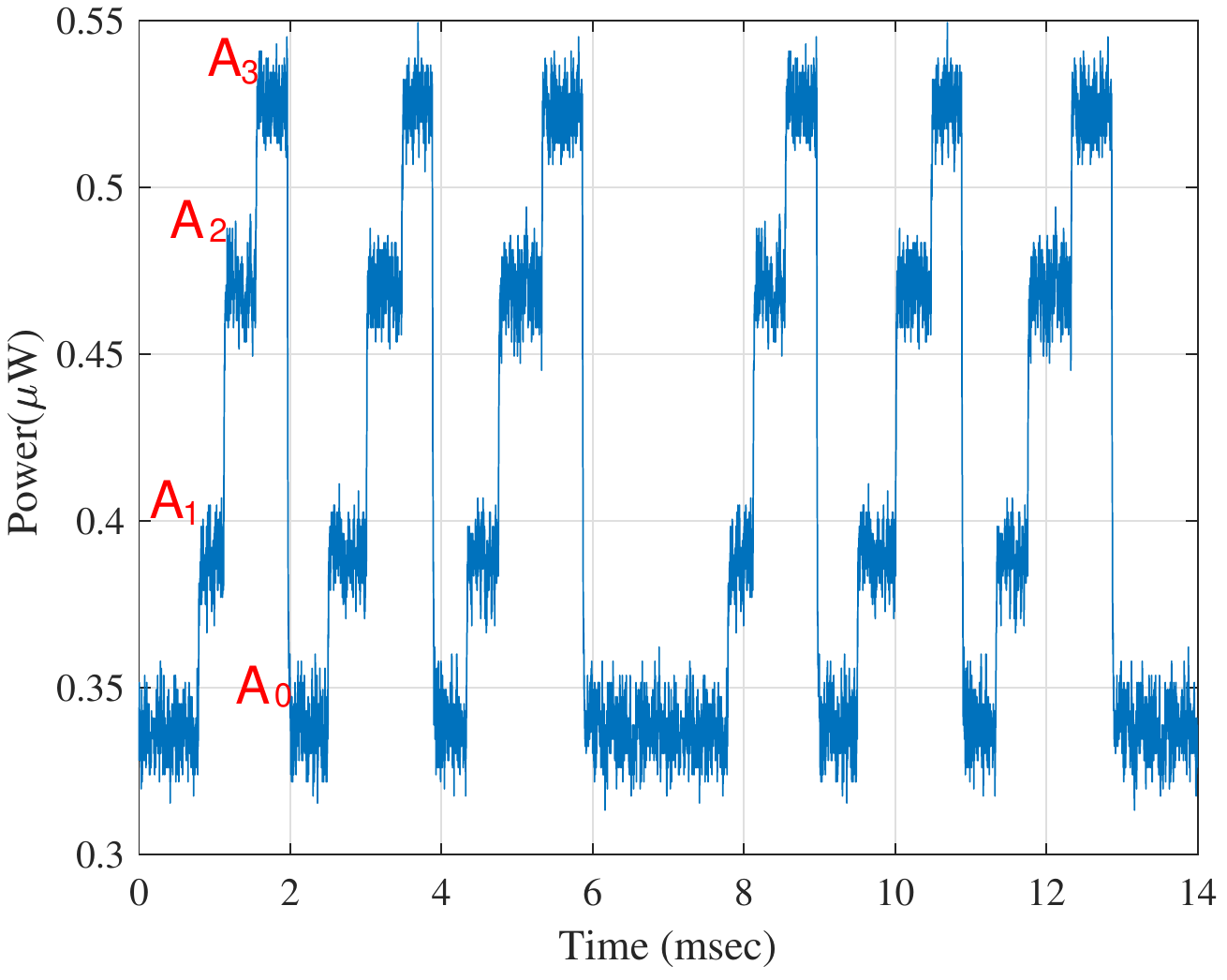}}\hfil
			
		\subfloat[]{\includegraphics[width=0.49\columnwidth]{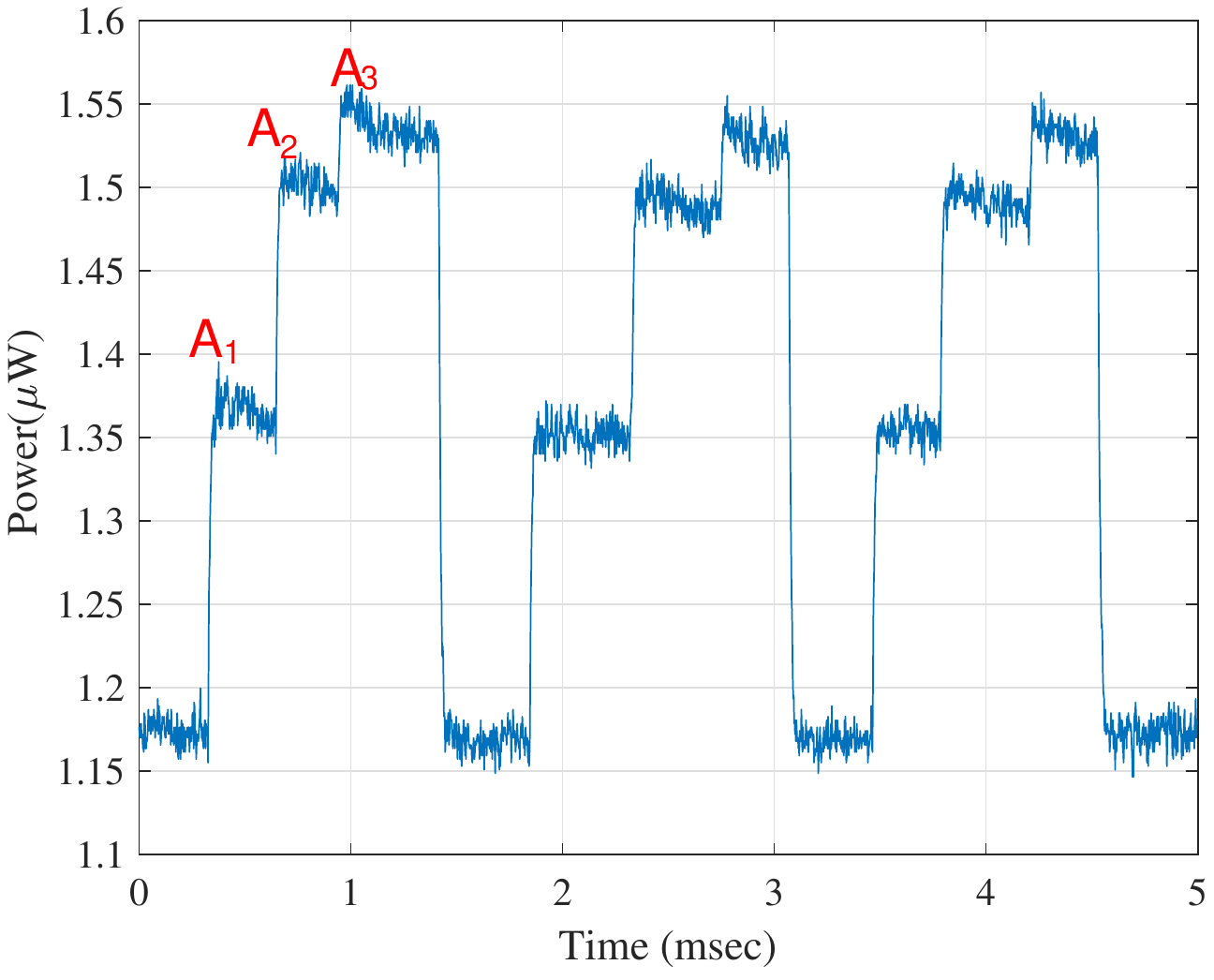}}	
			
	\caption{ASK modulation (a) Dell 3 LEDs, (b) Lenovo 3 LEDs, (c) Logitech 3 LEDs, (d) Silverline 3 LEDs.}
	\label{fig:ASKmodulation}
\end{figure}

Figure \ref{fig:ASKmodulation} shows four amplitude levels as measured from all of the keyboards when all three LEDs are in use. We employed four different states, starting with all three LEDs in the off state and sequentially turned the LEDs on until all of the LEDs were on (000, 100, 110, and 111). Note that we only distinguish between the number of LEDs turned on, as opposed to their location (e.g., the states 110, 011, and 101 represent the same amplitude). As can be seen in Figure \ref{fig:ASKmodulation}, we can distinguish between four different levels, when each amplitude level is modulated over 700$\mu$s, 500$\mu$s, 500$\mu$s, and 350$\mu$s. This implies the rate of approximately 1730, 2000, 2000 and 2850 different levels per second (3460, 4000, 4000, and 5710 bit/sec, respectively). 

\subsubsection{Transmission}
\begin{figure}[!ht]
\centering
		\subfloat[]{\includegraphics[width=0.49\columnwidth]{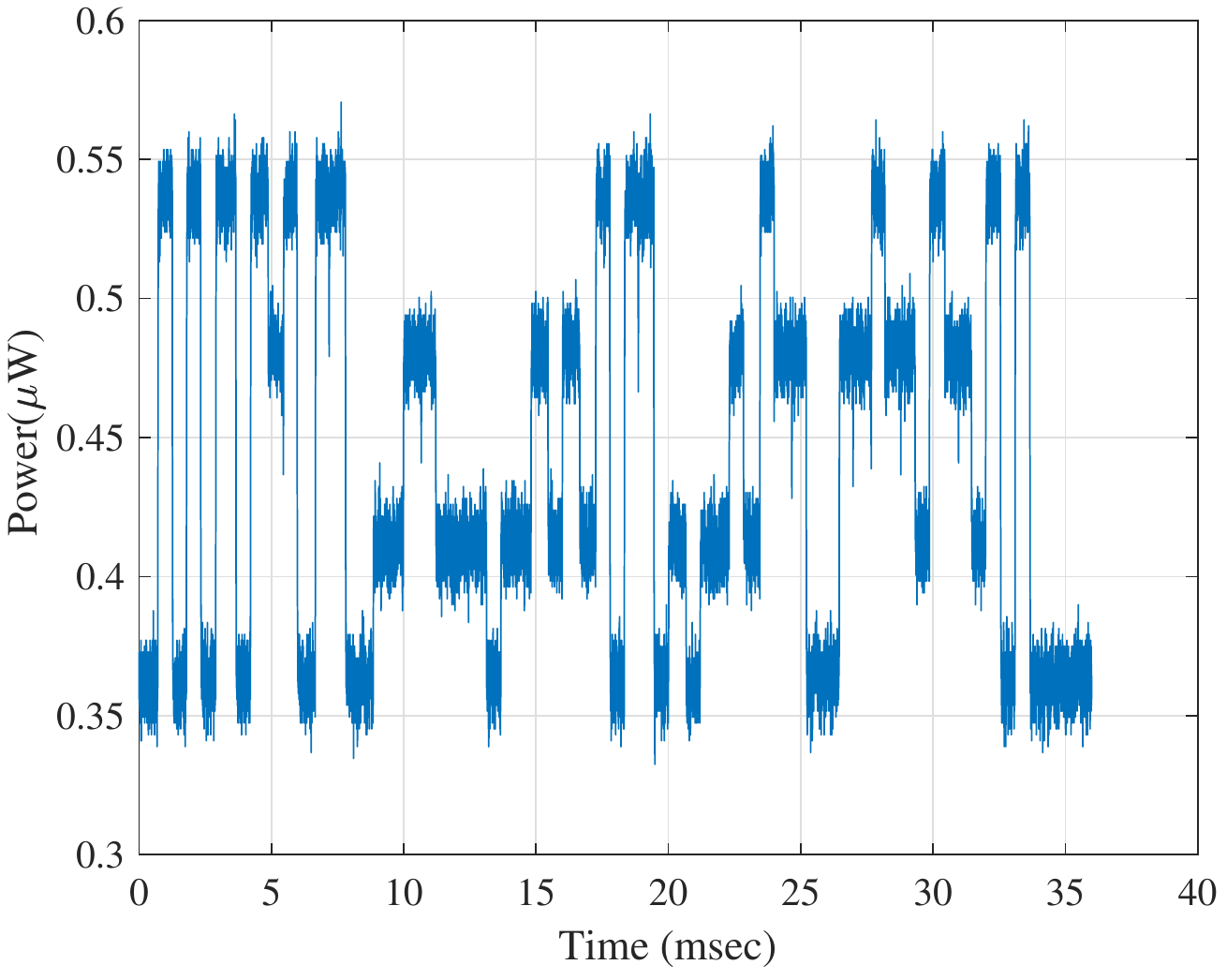}}\hfil
		\subfloat[]{\includegraphics[width=0.49\columnwidth]{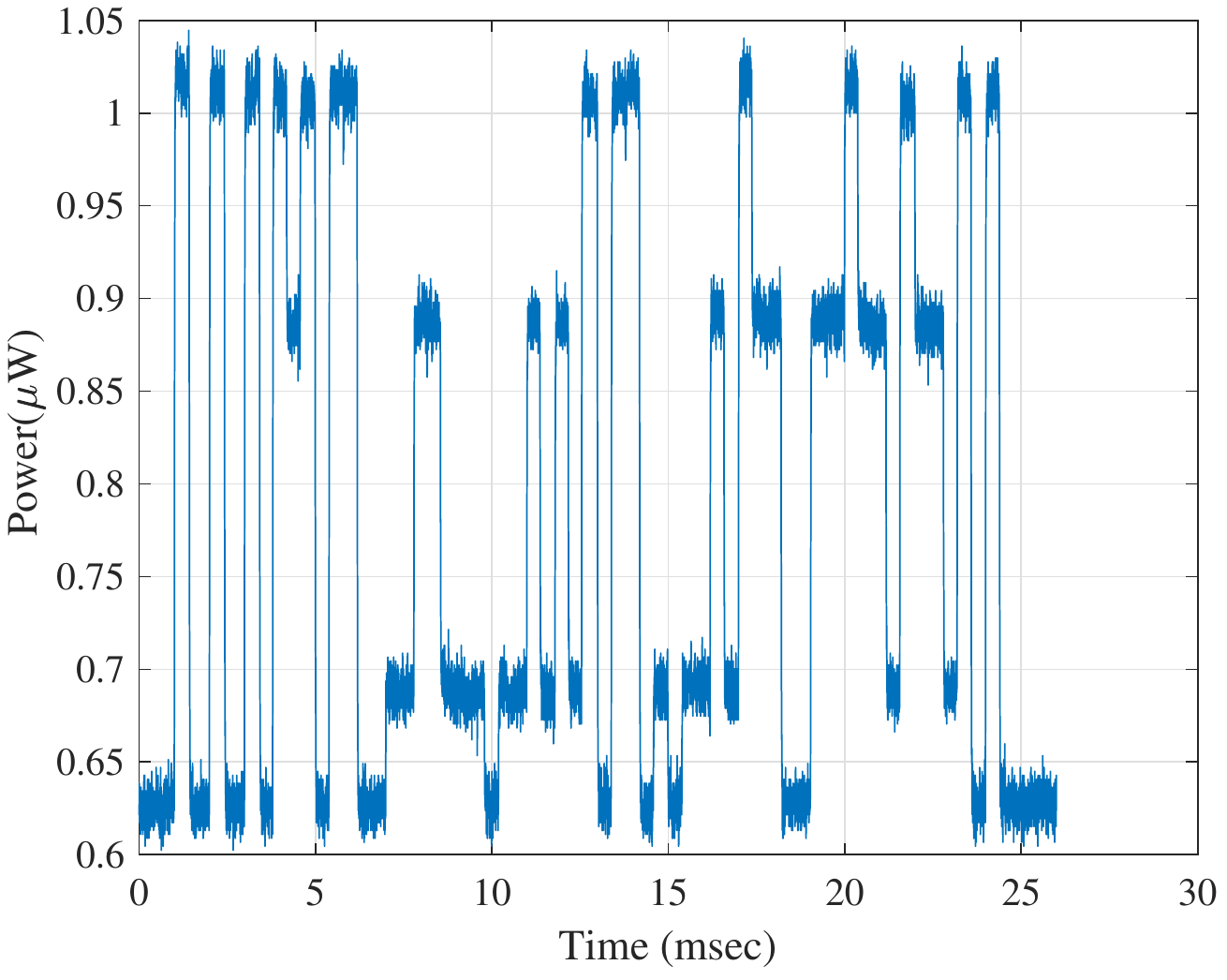}}
		
		\subfloat[]{\includegraphics[width=0.49\columnwidth]{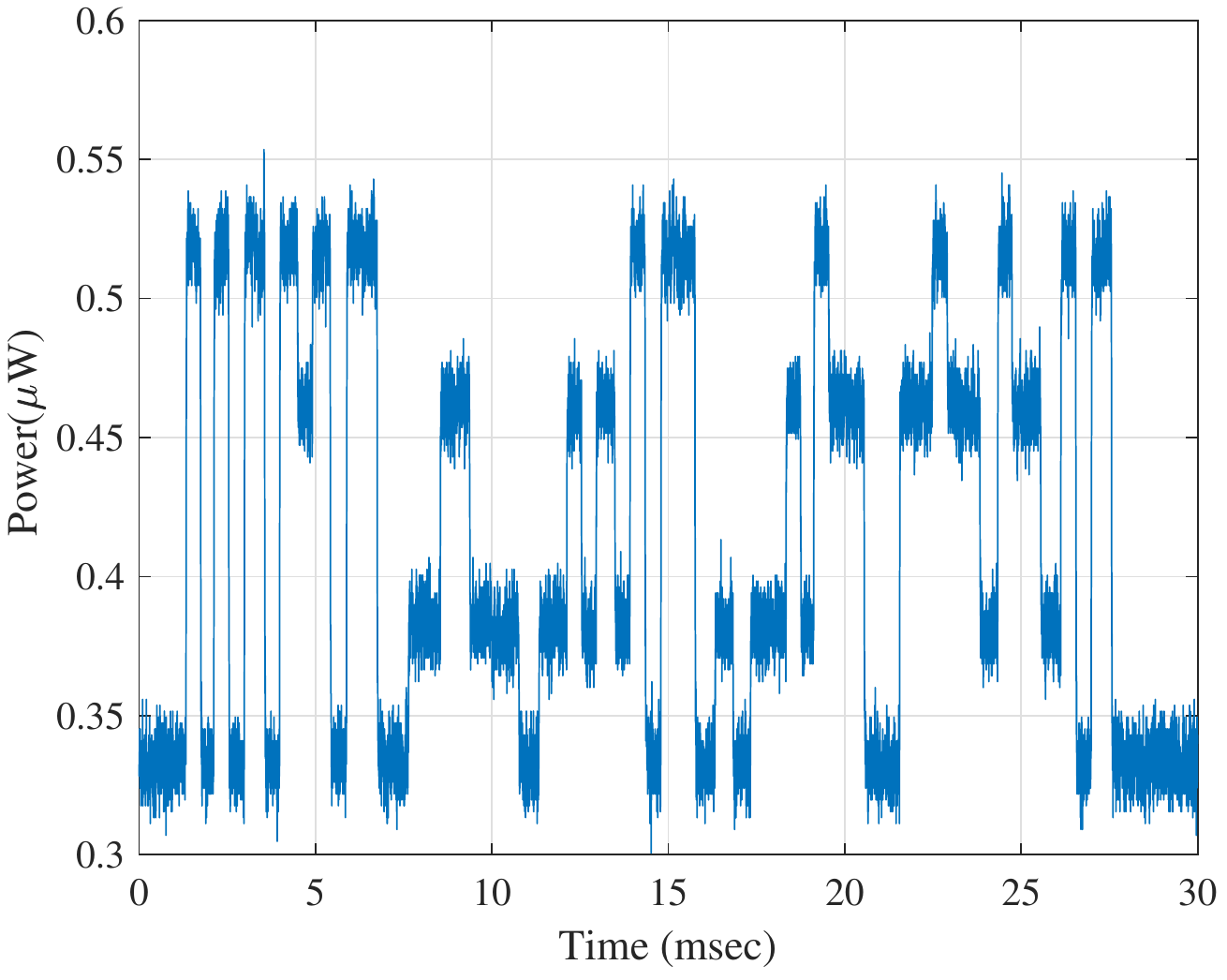}}\hfil
		\subfloat[]{\includegraphics[width=0.49\columnwidth]{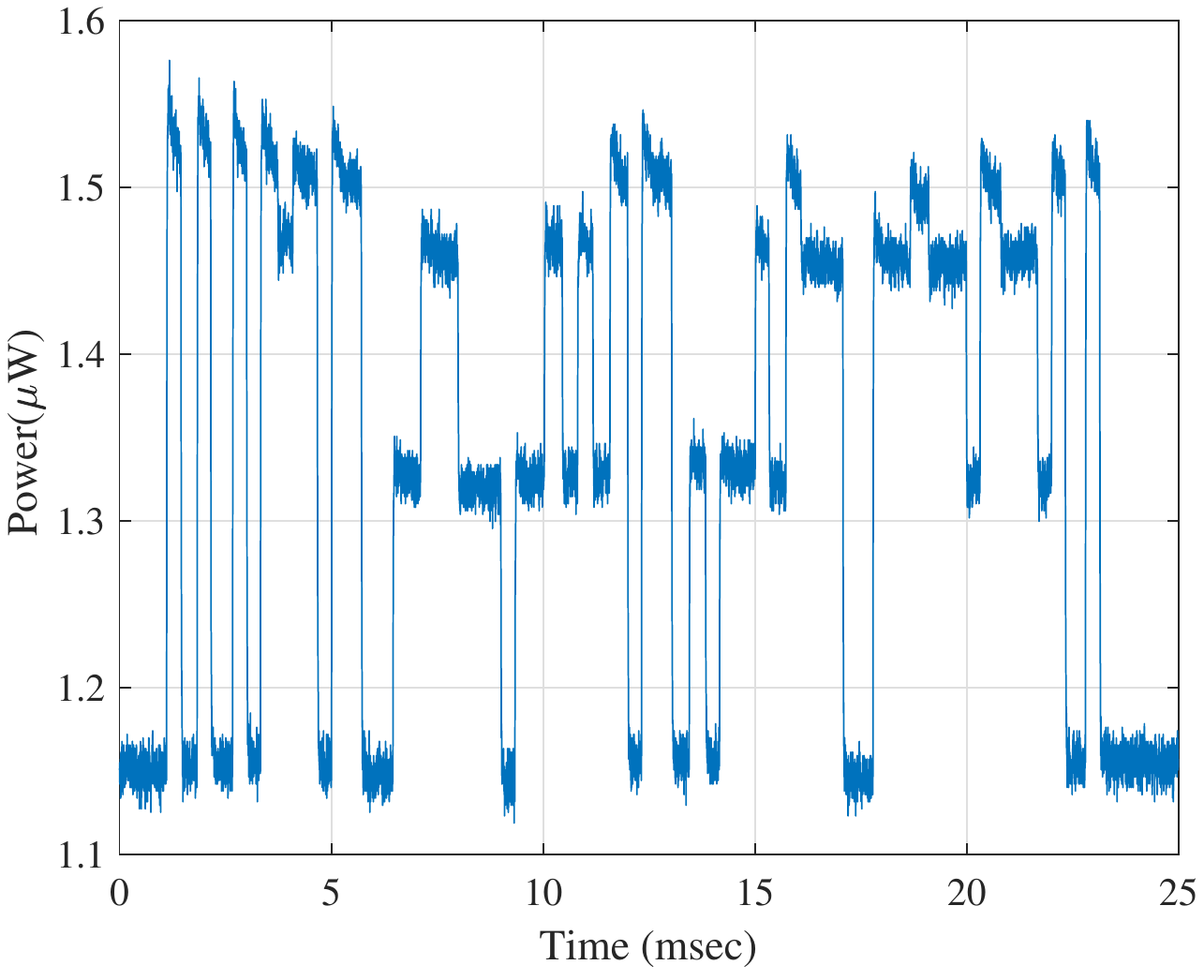}}	
	\caption{ASK data transmission (a) Dell 3 LEDs (b) Lenovo 3 LEDs (c) Logitech 3 LEDs  (d) Silverline 3 LEDs }
	\label{fig:data_transfer}
\end{figure}

Figure \ref{fig:data_transfer} shows the measurements in which a stream of bits was transmitted from all keyboards using ASK modulation via four LEDs. The stream was transmitted in 36ms, 25ms, 28ms, and 22ms which implies a bit rate of  1665, 2400, 2240 and 2725 bit/sec. Note that the bits are encoded with four amplitude levels $(A_0,$ $A_1,$ $A_2$ and $A_3)$. In this case, the measured BER was under 5\%.

The measured BER for the OOK and multiple LED modulation is provided in Table \ref{tab:BERall}.  

\begin{table}[]
	\caption{Bit Error Rates}
	\label{tab:BERall}
	\centering
	\renewcommand{\arraystretch}{1.3}	
			\begin{tabular}{c|c|c|c}
				\toprule
				\textbf{Keyboard} & \textbf{Modulation} & \textbf{ Bit-rate} & \textbf{BER in \%} \\
				\midrule
				      Dell        &         OOK         &    1666 bit/sec    &        3\%         \\
				\midrule
				      Dell        &    Multiple LEDs    &    3411 bit/sec    &       2.40\%       \\
				\midrule
				     Lenovo       &         OOK         &    2230 bit/sec    &       2.95\%       \\
				\midrule
				     Lenovo       &    Multiple LEDs    &    4640 bit/sec    &       6.70\%       \\
				\midrule
				    Logitech      &         OOK         &    2170 bit/sec    &       3.50\%       \\
				\midrule
				    Logitech      &    Multiple LEDs    &    4296 bit/sec    &       1.20\%       \\
				\midrule
				   Silverline     &         OOK         &    2697 bit/sec    &        8\%         \\
				\midrule
				   Silverline     &    Multiple LEDs    &    5155 bit/sec    &       3.10\%       \\
				\bottomrule
			\end{tabular}%
\end{table}